%% file: zhang.tex
\documentclass[12pt,psfig,aasms5]{aastex}
\def\fxu{{$\rm ergs\ cm^{-2}\ s^{-1}$}}

\def\lesssim{\,
\lower2truept\hbox{${<\atop\hbox{\raise4truept\hbox{$\sim$}}}$}\,}
\def\gtrsim{\,
\lower2truept\hbox{${>
\atop\hbox{\raise4truept\hbox{$\sim$}}}$}\,}

\received{} \accepted{}

\shortauthors{Zhang et al.} \shorttitle{Optical Candidates for
X-ray Sources}

\begin{document}

\baselineskip 24pt

\title{Multicolor Photometric Observations of Optical Candidates
to Faint ROSAT X-ray Sources in a 1 deg$^2$ field of the BATC
Survey}

\vskip 1cm

\author{\large
Haotong Zhang\altaffilmark{1}, Suijian Xue\altaffilmark{1},
David Burstein\altaffilmark{2}, \\
Xu Zhou\altaffilmark{1,3}, Zhaoji Jiang\altaffilmark{1}, Hong
Wu\altaffilmark{1}, Jun Ma\altaffilmark{1},
Jiansheng Chen\altaffilmark{1,3}, \\
and Zhenlong Zou\altaffilmark{1} }

\vskip 1.5cm \affil{\large \altaffiltext{1}{National Astronomical
Observatories, Chinese Academy of Sciences, A20 Datun Rd.,
Chaoyang District, Beijing 100012} \altaffiltext{2}{Department of
Physics and Astronomy, Box 871504, Arizona State University,
Tempe, AZ 85287--1504} \altaffiltext{3}{Beijing Astrophysics
Center \& Department of Astronomy, Peking Univ., Beijing 100871,
China} }

\email{zht@lamost.org}

\clearpage
\begin{abstract}

We present optical candidates for 75 X-ray sources in a $\sim 1$
deg$^2$ overlapping region with the medium deep ROSAT survey
(Molthagen et al. 1997). These candidates are selected using the
multi-color CCD imaging observations made for the T329 field of
the Beijing-Arizona-Taipei-Connecticut (BATC) Sky Survey, which
utilizes the NAOC 0.6/0.9m Schmidt telescope with 15
intermediate-band filters covering the wavelength range 3360-9745
\AA. These X-ray sources are relatively faint (CR $<< 0.2 s^{-1}$)
and thus mostly are not included in the RBS catalog, they also
remain as X-ray sources without optical candidates in a previous
identification program carried out by the Hamburg Quasar Survey.
Within their position-error circles, almost all the X-ray sources
are observed to have one or more spatially associated optical
candidates within them down to the magnitude $m_V \sim 23.1$.  We
have classified 149 of 156 detected optical candidates with 73 of
the 75 X-ray sources with a new method which predicts a redshift
for non-stellar objects, which we have termed the SED-based Object
Classification Approach (SOCA).  These optical candidates include:
31 QSOs, 39 stars, 37 starburst galaxies, 42 galaxies, and 7
``just'' visible objects.  Twenty-eight X-ray error circles have
only one visible object in them: 9 QSOs, 3 normal galaxies, 8
starburst galaxies, 6 stars, and two of the ``just'' visible
objects.  We have also cross-correlated the positions of these
optical objects with NED, the FIRST radio source catalog and the
2MASS catalog.  Separately, we have also SED-classified the
remaining 6011 objects in our field of view. Optical objects are
found at the $6.5\sigma$ level above what one would expect from a
random distribution, only QSOs are over-represented in these error
circles at greater than 4$\sigma$ frequency.  We estimate
redshifts for all extragalactic objects, and find a good
correspondence of our predicted redshift with the measured
redshift (a mean error of 0.04 in $\Delta z$.  There appears to be
a supercluster at z $\sim$ 0.3-0.35 in this direction, including
many of the galaxies in the X-ray error circles are found in this
redshift range.

\end{abstract}

{\bf keywords:} X-rays: galaxies - galaxies: active - catalog:
surveys

\clearpage

\section{Introduction}

Combined optical and X-ray data allow one to obtain information
about the luminosity functions of various types of X-ray sources
as well as their evolution with redshift. In turn, this
information can be used to further constrain models for the
production of the X-ray background at different flux levels.  Much
effort has so far been made on the optical identification of the
X-ray sources in the ROSAT/Bright Source (RBS) catalog (e.g.,
Voges et al. 1999; Rutledge et al. 2000) as well as of X-ray
sources in some individual ROSAT deep survey observations (e.g.,
Lehmann et al. 2001).

Yet, it is often unknown how much of the detections that occur in
X-ray error circles are real associations of optical counterparts
to these X-ray sources, and how much of these associations are due
to random chance.  To do this, one needs to have detected and
identified all optical objects in a given image, and then see what
percentages of these objects (QSOs, galaxies, stars) are then
found near or within the areas covered by the X-ray error circles.
This is precisely the kind of data we have for 75 X-ray sources
detected with the ROSAT PSPC to a flux limit $S_x({\rm 0.1-2.4
keV}) \geq 5.3\times10^{-14}$\ \fxu, in a 1 deg$^2$ field of view,
as this field of view was also observed in our multicolor images
for the Beijing-Arizona-Taipei-Connecticut Sky Survey (BATC
survey).  The relevant X-ray and optical data are presented in
\S~2.  Details of the object classification procedure as well as
selection of the X-ray candidates are given in \S~3. Associated
information that can be gleaned from these data are given in \S~4.
We summarize our results in \S~5.

\section{The data and analysis}
\subsection{The X-ray data}

The X-ray data come from a catalog obtained from a medium deep
ROSAT survey in the HQS field HS 47.5/22 (Molthagen, Wendker, \&
Briel, 1997). The survey consists of 48 overlapping ROSAT PSPC
pointings which were added up to produce a final catalog
containing 574 X-ray sources with broad band (0.1-2.4 keV) count
rates between $\sim3\times10^{-3}\rm cts\ s^{-1}$ and $\sim0.2\rm
cts\ s^{-1}$, in a field of view (FOV) of $\sim 2.3\rm\ deg^2$.
Molthagen et al. adopt an X-ray error circle of 2$\sigma$ + 10$''$
in radius, with the value of $\sigma$ coming from their
observations.  This is the X-ray error circle used in the present
analysis.

\begin{figure}
\plotone{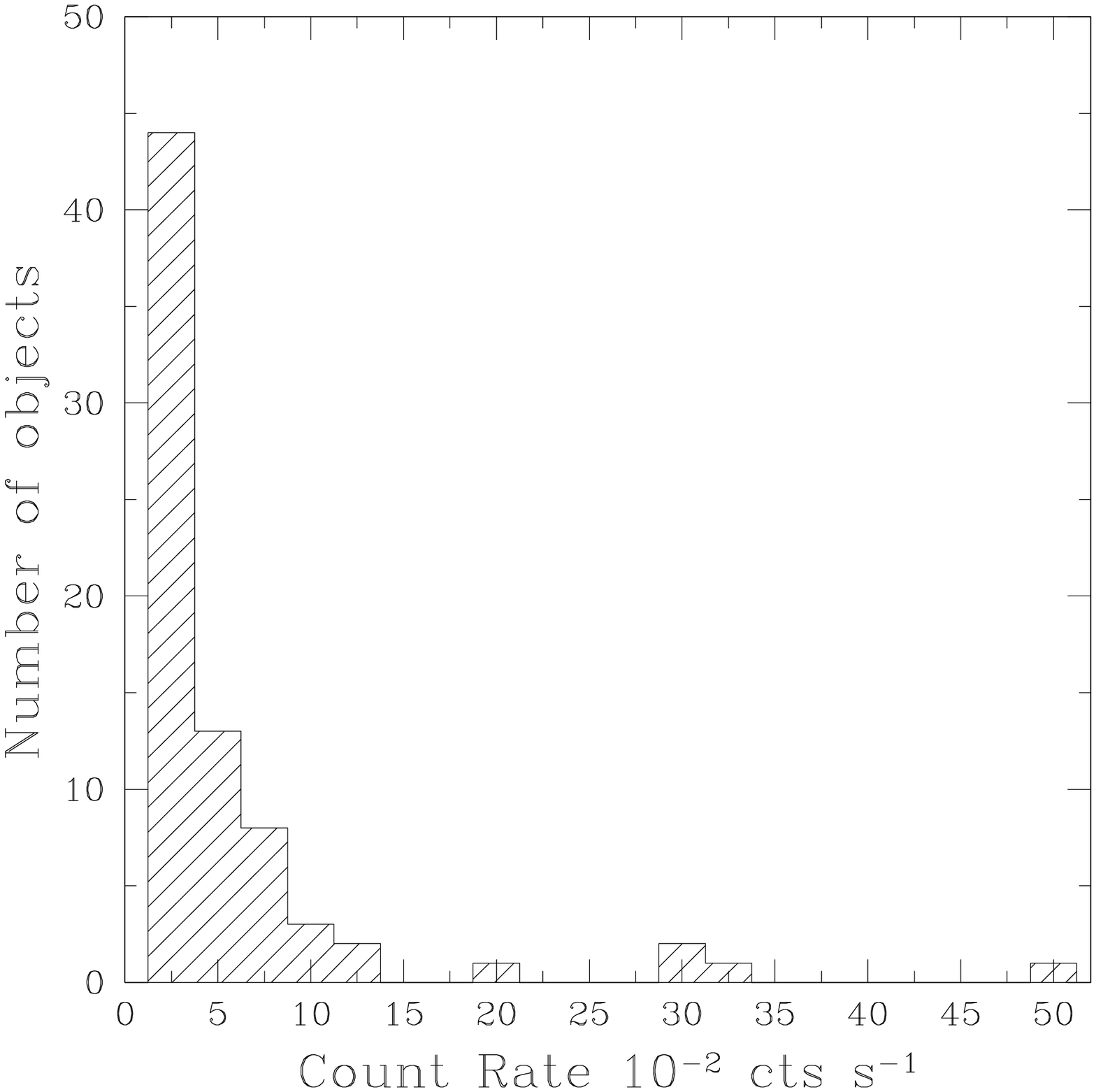} \caption{ The brightness distribution in count
rate (0.1-2.4 keV) of the 75 X-ray sources from a medium deep
ROSAT survey.} \label{f1}
\end{figure}

\begin{figure}
\plotone{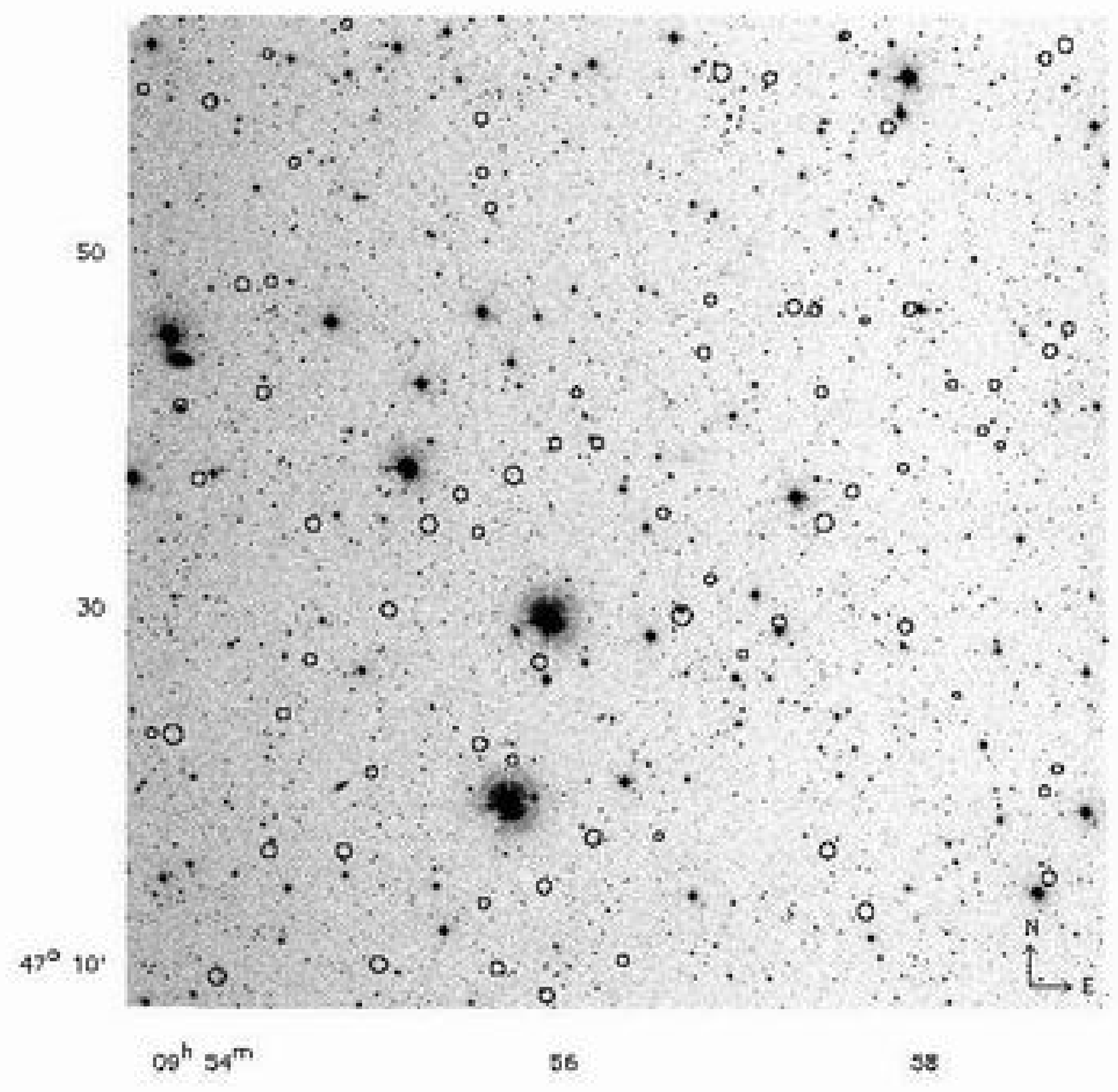} \caption{Distribution of X-ray sources in BATC
optical image, X-ray error circles are show as circles with
corresponding radius.} \label{x_dist}
\end{figure}

There was a preliminary identification of these X-ray sources with
the HQS plates (Molthagen et al, 1997). Only a few objects, all
brighter than $m_B\approx18^m.5$, have recognizable spectra. At
$m_{\rm B}>18^m.5$, many objects are generally classified as weak
and extremely blue, blue or red. For many X-ray sources no
spectral classification was possible, the optical object simply
being visible or the field of view empty.  75 of the 574 HQS
sources fall on one program field of the BATC survey, T329,
centered at 09:56:24.46, +47:35:08.4 (J2000), forming a subsample
of the ROSAT medium deep survey in a 1 deg$^2$ field. (One-third
of the BATC fields are located with a known quasar in its center.
For field T329, this quasar is PC0953+4749 with z = 4.46,
originally discovered by Schneider, Schmidt \& Gunn 1991.
Ironically, this QSO is not an X-ray source in the HQS field.) The
X-ray brightness distribution of these 75 sources is shown in
Fig.\ref{f1}.  The distribution of these 75 sources in our field
of view in shown in Fig.\ref{x_dist}.

Molthagen et al. associate 25 optical candidates for these 75
X-ray sources, or a frequency of 1/3: 6 QSOs or active galaxies; 7
QSO/active galaxy candidates (classified as such or extremely
blue); 1 star; 8 stellar candidates; 1 galaxy candidate; 2 faint
red objects; 5 unidentified spectra (including overlaps); 39
visible on the HQS direct plate only; and 6 empty fields (i.e., no
counterpart on the HQS plate).

\subsection{The BATC optical data}

Optical observations of BATC field T329 were carried out from
1996-1999 as part of the BATC Survey.  Our survey utilizes the
0.6/0.9m Schmidt telescope of the Xinglong Observing Station of
the National Astronomical Observatory of China (NAOC), equipped
with 15 intermediate-band filters covering the wavelength range
3360-9745\AA. With this facility our survey is designed to do
multi-color CCD ($2048\times2048$) imaging of 500 selected, $\sim
1$ deg$^2$ fields-of-view for multiple scientific purposes (cf.
Fan et al. 1996; Shang et al. 1998; Zheng et al. 1999; Zhou et al.
1999; Yan et al. 2000; Kong et al. 2000; Ma et al. 2002; Wu et al.
2002).

The dataset for T329 consists of a number of individual direct CCD
images in each of the 15 BATC passbands.  These images are first
treated individually (bias, dark and flat-fielding corrections)
and then combined to comprise deep images.  Information on the
passbands used for the present study, including filter parameters,
total exposure time, number of flux calibration images obtained,
and the magnitude limit for that passband are given in
Table~\ref{table1}. Details on the BATC flux calibration procedure
are given in several previous papers (Fan et al. 1996; Zhou et al.
1999; Yan et al. 2000) and the reader is referred to those papers
for this information. Further discussion of the observations made
in field T329 that are separate from the X-ray identification
issue are given in Zhou et al. (2003).  The final product of the
BATC observations of field T329 is a catalog of 6160 point-like
optical objects in our 58 arcmin$^2$ field of view, with
astrometry and photometry in 15 colors.

\subsection{SED classification}

We are in the process of developing a SED-based Object
Classification Approach (termed SOCA) for the BATC photometric
system (Zhang, et al., in preparation). The SED of each object in
our field of view, observed through $n$ filters, is compared to
the SED computed for a set of template spectra. The aim is to find
the best fit between the observed photometry and the model
photometry through a standard $\chi^2$ minimization procedure:

$$ \chi^2=\sum^N_{i=1}(\frac{f^{obs}_i-A \cdot
f^{temp}_i}{{\sigma}^{obs}_i})^2 $$

where $f^{obs}_i$ and $f^{temp}_i$ are the observed and the
template fluxes in the $i$th band respectively, ${\sigma}^{obs}_i$
is the error on the observed flux in this band, $A$ is the
normalization constant can be calculated by minimize the $\chi^2$.
In order to apply this method to SED classification, we currently
employ three sets of template spectra:

\begin{enumerate}

\item The stellar library of Gunn \& Stryker (1983) is used
including most spectral classes on the MK system (this will be
updated when new data are available);

\item The observed spectra of nearby galaxies (Kinny et al.
1996),including normal galaxies (Elliptical ,S0, Sa, Sb and Sc)
and starburst (SB) galaxies (SB1-6) with different internal
extinctions are used.  Normal galaxies are redshifted from 0 to
1.0 in step of 0.01, SB galaxies are redshifted from 0 to 1.5 in
step of 0.01.

\item A QSO template set is composed of series of simulated quasar
spectra. These spectra have been constructed by fixing the
emission line intensity ratio (cf. Wilkes 1986), while varying the
$ly\alpha$ equivalent width (65 $\pm$ 34 \AA) and the continuum
index $\alpha$ (-0.75$\pm$ 0.5).  $Ly\alpha$ forest absorption has
been modeled according to M{\o}ller (1990) and Madau (1995).
Redshift estimates are set between 0.0 and 6.0 in steps of 0.01 in
$z$.

\end{enumerate}

Representative template spectra used in the present paper are
given in Fig.\ref{f2}. The template SEDs are obtained by
convolving the template spectra with the measured passband of each
filter. As the template SEDs are morphologically-classified, some
templates may represent two or more morphologically-similar
classes. For example, an SED classified as a starburst galaxy can
also possibly be matched to that of a QSO.

A value of $\chi^2$ is calculated for the correspondence of every
template to each object SED.  The minimum $\chi^2$ for each kind
of template (star/galaxy/QSO/starburst galaxy) is calculated. The
template with the $\chi^2$ minimum fit is taken as the best fit.
In this fitting process we include those objects with at least 5
filter observations (such as saturated stars).  The redshift
estimates found for non-stellar objects (galaxies, QSOs) by this
template-fitting process are useful for statistical studies of
this field of view.

\section{Optical Candidates}

\subsection{Optical Objects near or within the X-Ray Error Circles}

The CCD limiting magnitudes range from 20.5 to 23.5 mag, tending
to be fainter in the bluer filters, and brighter in the more
sky-limited redder filters (cf. Table~\ref{table1}).  Our deeper,
direct CCD observations, combined with our ability to classify the
SEDs of the objects detected, permit us both to detect more
objects than the HQS survey, as well as to classify more of the
objects detected.

The total area covered by the X-ray searches we have done for the
75 X-ray sources corresponds to 31.52 arcmin$^2$ or 0.00937 of the
3364 arcmin$^2$ sky area subtended by the BATC CCD.  This area
includes additional area searched beyond the nominal 2$\sigma$
error circle for 13 (17\%) of the X-ray-detected objects, most of
these within 1-2$''$ of the original error circle.  The 6160
objects detected in the full image field were selected with the
same criteria as those we use for the X-ray error circle. If the
optical objects and the X-ray sources are randomly associated, we
expect to detect $31.52/3364 \times 6160 = 58$ optical objects.

Our observations find optical candidates (stars, galaxies, galaxy
groups, starburst galaxies, QSOs) in 73 of the 75 X-ray error
circles. We detect a total of 156 optical objects in these 75
X-ray error circles. Of these we can definitely SOCA-identify 140,
tentatively identify 9 more (7 galaxies and two stars),  7 objects
are only ``visible,'' and one X-ray circle (RXJ0955.5+4735) which
is blank in our image (but two stars are just outside this error
circle).  This makes a total of 149 optically-identified
candidates found in the BATC catalog that can be SOCA-classified
and are found in or near 73 of the 75 X-ray error circles in our
field of view. One of the two remaining X-ray sources
(RXJ0954.0+4756) has a ``just'' visible object within the X-ray
error circle that has a position coincident with a known radio
source, but is too faint to obtain a reliable SED.  The other
remaining source (RXJ0953.7+4722) also has one ``just'' visible
source within its error circle.

We have a difference in objects detected in the X-ray circles to
those randomly expected of $149 - 58 = 91\pm14$(assuming Gaussian
errors). The difference between detected objects and random
placement of objects in these X-ray error circles is significant
at the 91/14 = 6.5$\sigma$ level. It would appear that the X-ray
circles do tend to include more objects than randomly placed
circles put on the rest of this field of view, when the data are
sampled to faint magnitude levels.

Table~\ref{table2} gives the relevant information on the optical
candidates that are associated with these X-ray sources.  The
first 4 columns in Table~\ref{table2} come from the original X-ray
catalog: X-ray source name, brightness in the 0.1--2.4 keV
passband, 2$\sigma$ error circle radius in units of arcseconds,
and the original HQS identification. The label assigned to each
candidate optical object, a,b,c,$\ldots$, plus the observed
position of the optical candidate (in J2000 coordinates) are given
in the next three columns.  Columns 8-12 give the derived
information for each optical candidate: $\Delta r$ is the offset
of the optical candidate position from the center of the X-ray
error circle; $m_V$ denotes the V magnitude of the optical
candidate (an upper limit is given if the candidate is only
visible, but not measurable), derived from the relation: $m_{\rm
V}=m_{\rm g} + 0.3233(m_{\rm f} - m_{\rm h}) + 0.0590$ (Zhou et
al., 2003); $f_{xo}$ is the ratio of X-ray to optical flux,
calculated from the 0.1-2.4 keV count-rate and optical V
magnitude, vis. $f_{\rm xo} = {\rm log}(f_x/f_o) = {\rm log(PSPC\
counts/s}\times10^{-11}) +0.4m_V + 5.37$ (Maccacaro et al. 1988);
Pred z is the redshift that the SOCA estimates for galaxies and
QSOs; Where there are known candidates that are clearly identified
on our images, the identity of these candidates are given. Finding
charts of all the X-ray sources in our summed image at 7050\AA
(our j filter) are given in Fig.\ref{idt}.

The SEDs for all 149 SOCA-identified optical candidates in or near
73 of the 75 X-ray source error circles are given in
Fig.~\ref{sed}, in which also the best template fit is plotted for
each optical candidate.  The label on each SED gives the template
plotted, the predicted redshift (if galaxy or QSO), and the value
of the $\chi^2$ fit.  In the case of two known objects (a nearby
galaxy and a bright star; see next section), their
previously-known identifications are given in place of the SOCA
classification in Table~\ref{table2}.

\subsection{Candidate Associations}

Most X-ray sources contain more than one optical candidate within
their error circles.  Choosing which one is the probable X-ray
source is educated guesswork at best. Rather than assign
probabilities of the likelihood of each optical candidate's
association with these X-ray sources, we prefer to give the reader
the statistics of how the candidates relate to the full data on
the 6160 objects found in our field of view.

In the 75 X-ray error circles, to a magnitude limit of V $\approx$
23, we have found: 31 QSOs, 39 stars, 37 starburst galaxies, 42
galaxies, and 7 ``only visible'' objects.  If we take the 6160
objects we find in our field of view, excluding the objects within
the X-ray error circles, the analogous counts for these objects
are: 341 QSOs, 1912 stars, 1508 starburst galaxies, 2076 galaxies,
and 174 objects unclassified for a variety of reasons (too few
filters observed, in the halo of a bright star, etc.).  If we
assume these objects are randomly distributed in this field of
view, we expect random associations within our X-ray error circles
to be 3.2 QSOs, 17.9 stars, 14.1 starburst galaxies, and 19.5
galaxies (discarding the unclassified objects).  Therefore, the
objects found in the X-ray error circles rather than the random
placement of those identified objects in the field of view are:
QSOs: $31-3.2 =27.8\pm5.8$; stars: $39-17.9 = 21.1\pm7.9$ ;
starburst galaxies: $37-14.1 =22.9\pm7.1$; and galaxies: $42-19.5
= 22.5\pm7.8$.

\begin{figure}
\plotone{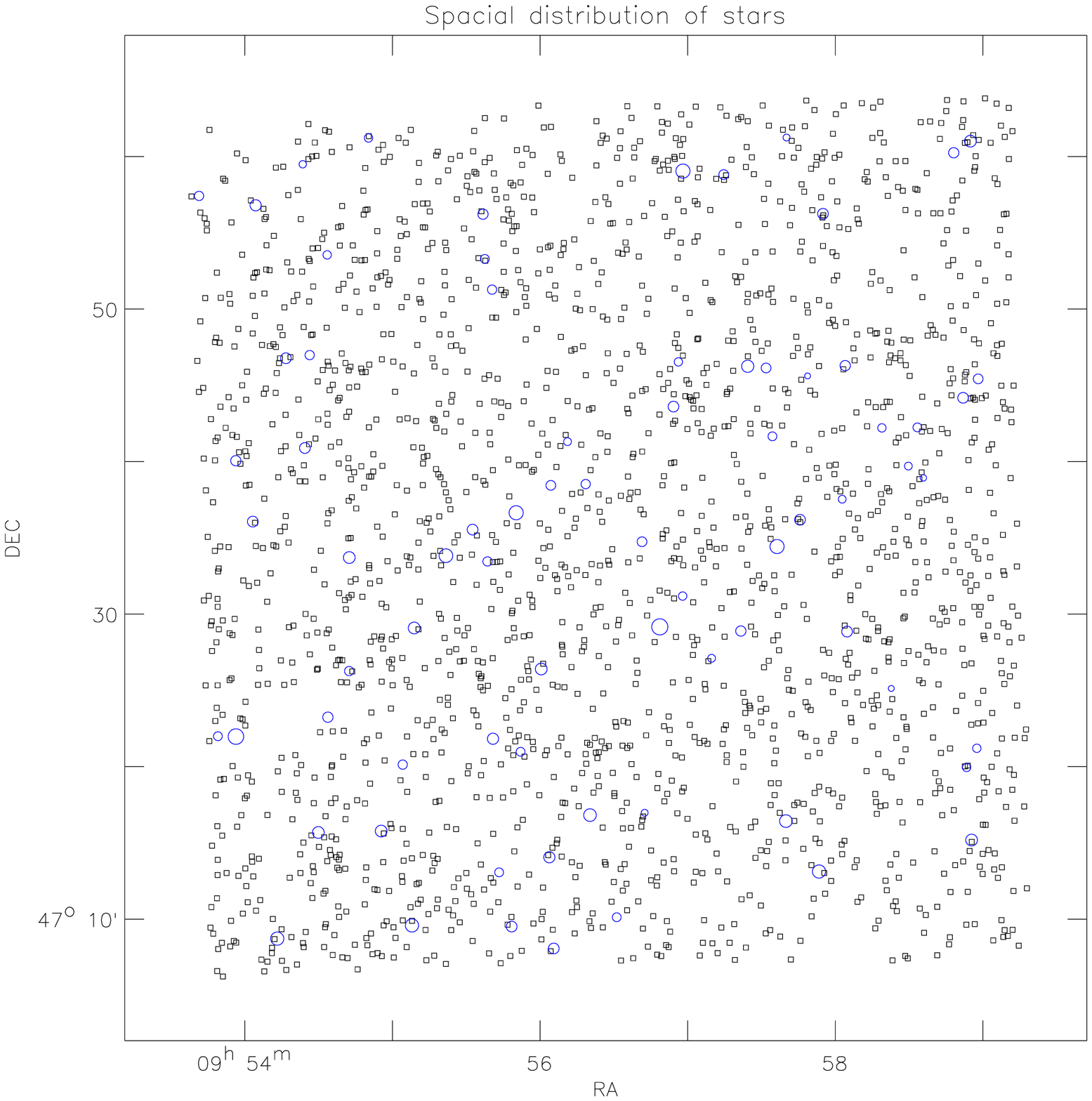}
\caption{Spacial Distribution of star candidates
relative to the X-ray error circles in T329 field.}
%\figurenum{7a}
 \label{stardis}
\end{figure}

\begin{figure}
\plotone{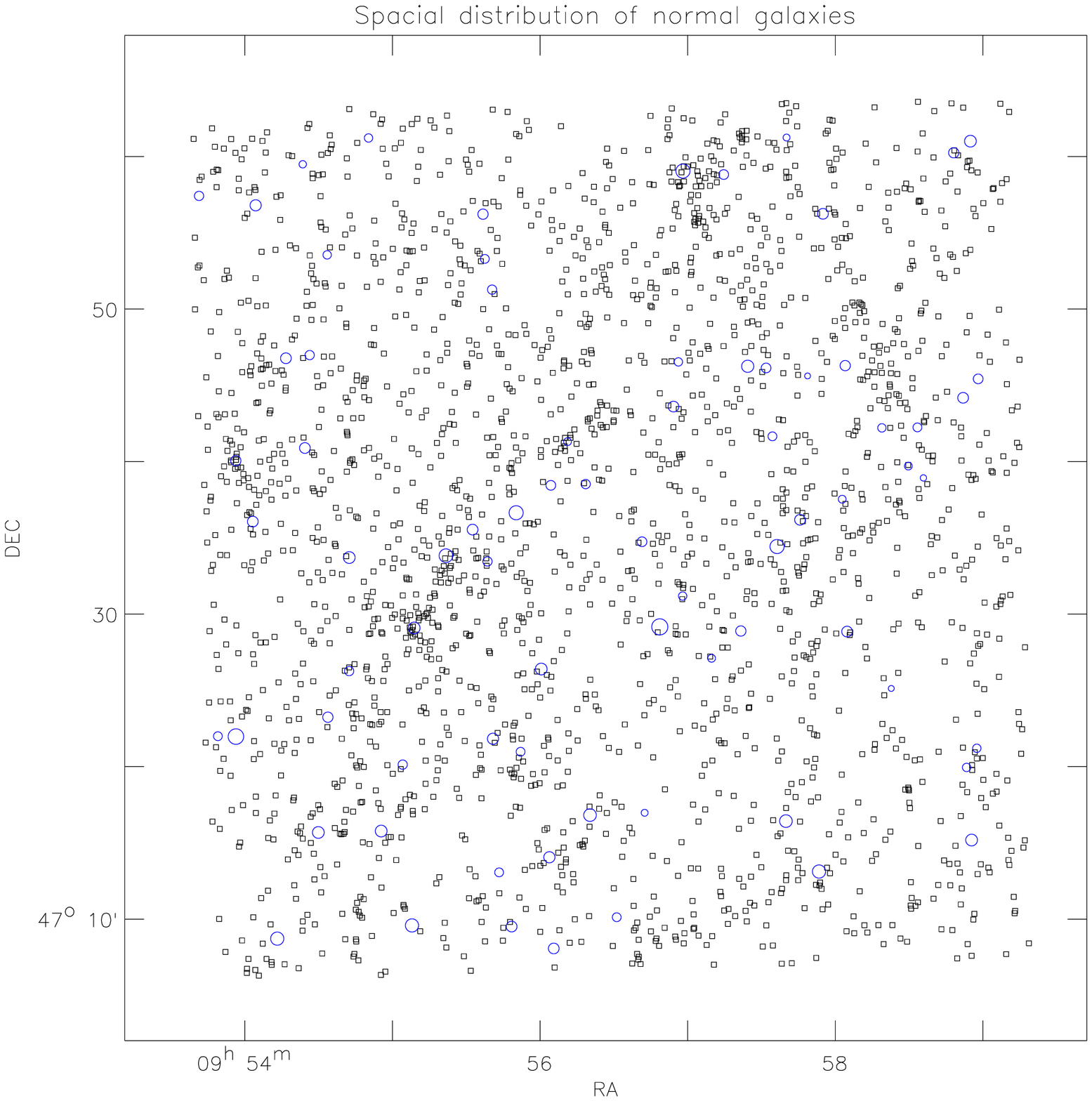}\caption{ Spacial Distribution of normal galaxy
candidates relative to the X-ray error circles in T329 field.}
%\figurenum{7b}
 \label{galdis}
\end{figure}

\begin{figure}
\plotone{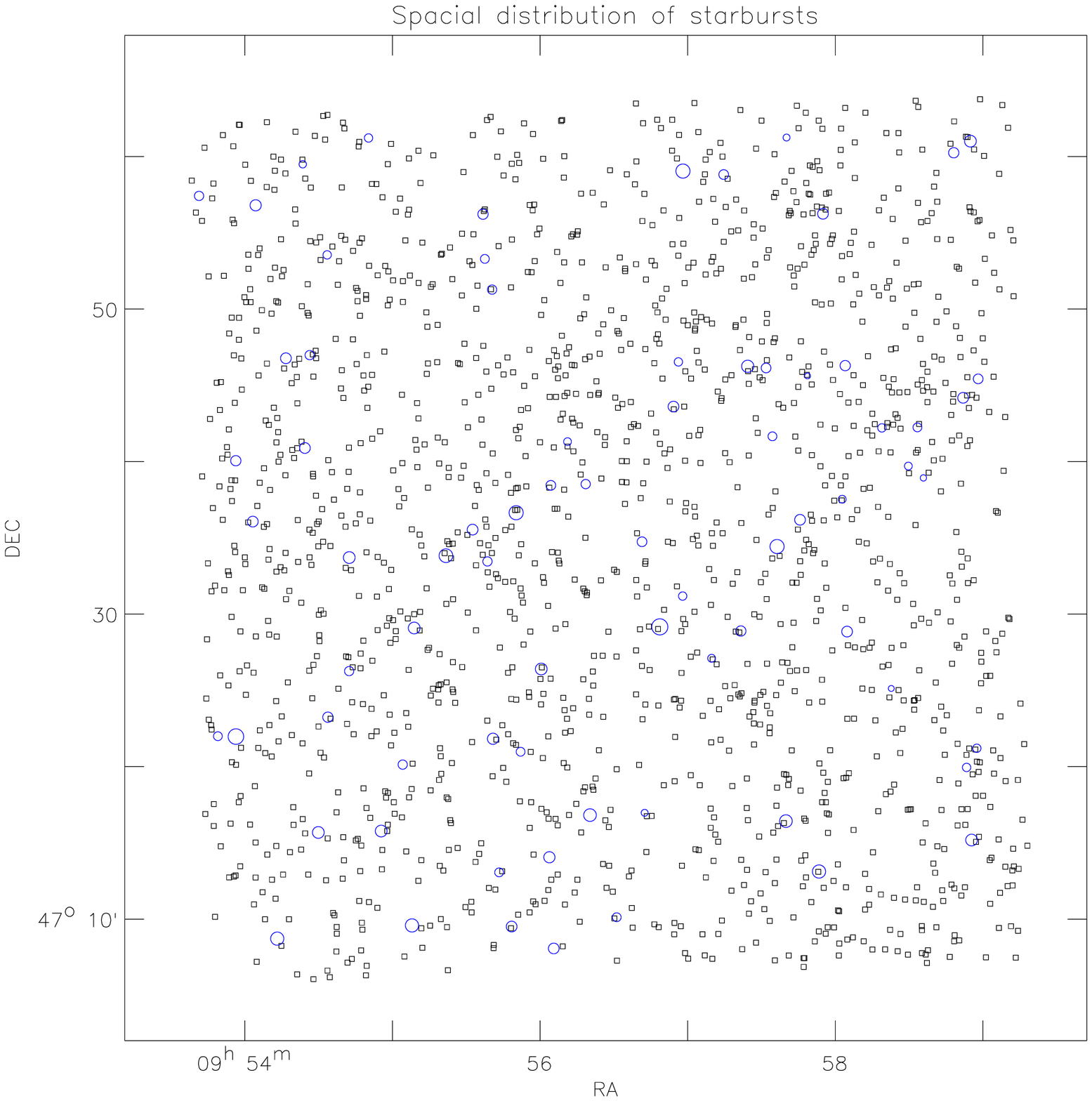}\caption{Spacial Distribution of starburst galaxy
candidates relative to the X-ray error circles in T329 field.}
%\figurenum{7c}
 \label{sbdis}
\end{figure}

\begin{figure}
%\epsscale{0.1}

\plotone{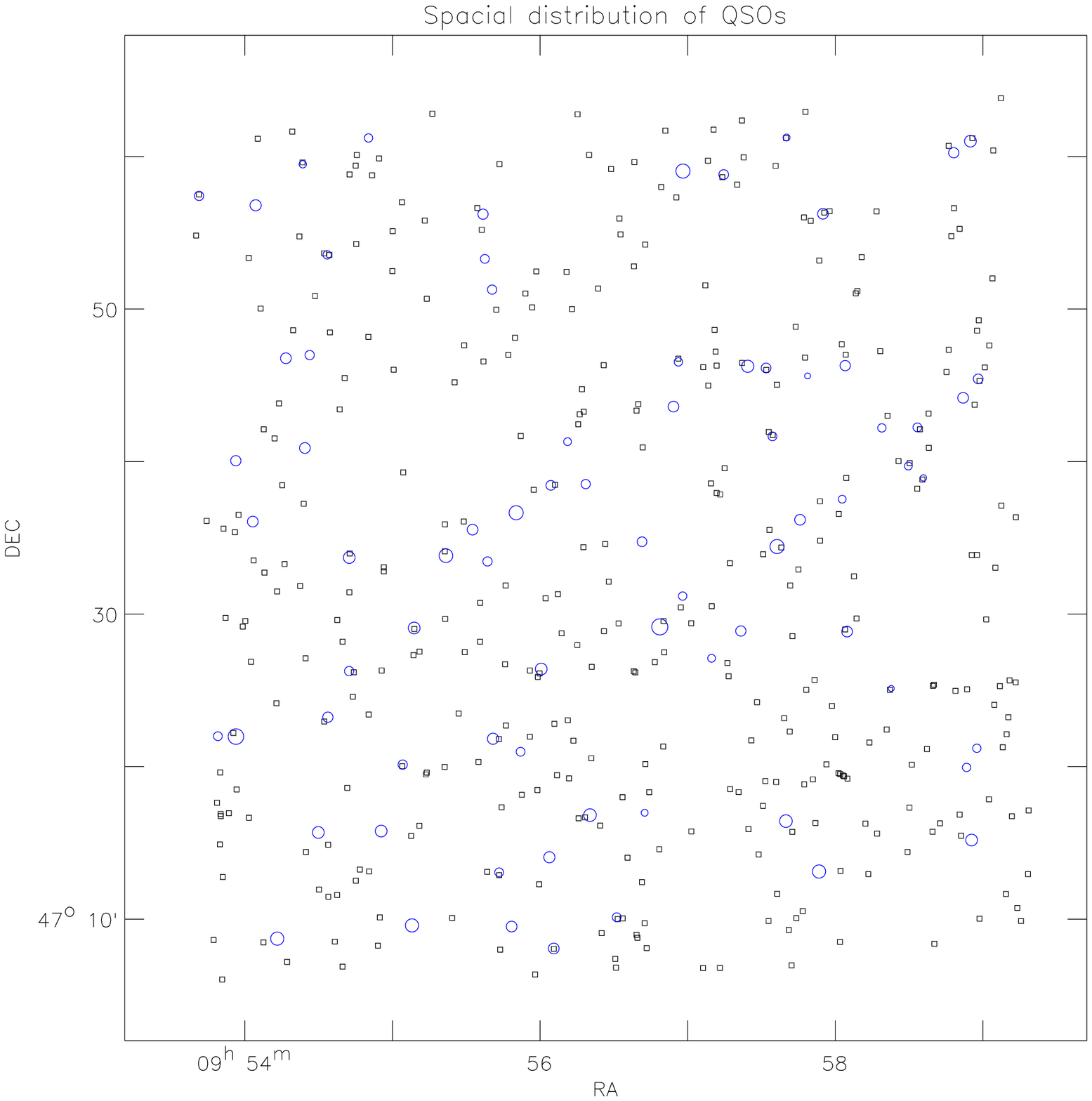}\caption{Spacial Distribution of QSO candidates
relative to the X-ray error circles in T329 field.}
%\figurenum{7d}
 \label{qsodis}
\end{figure}

On the plus side, all objects are more represented within the
X-ray error circles that what would be randomly there. On the
minus side, only the QSOs have a highly significant overdensity
($4.7\sigma$) within the X-ray error circles, while stars,
starburst galaxies and galaxies are only there at the
2.8-3.2$\sigma$ level.  In Fig\ref{stardis}-\ref{qsodis} we show
the distribution of each kind of object (QSO, star, starburst
galaxy, galaxy) in our field of view, relative to the placement
and size of the X-ray error circles. That QSOs are statistically
most reliable detections comes as little surprise, as this was
already well-known (e.g., Shanks et al. 1991, Georgantopoulos et
al. 1996, McHardy et al. 1998).

We only find one optical candidate in 28 of these error circles,
these being: 9 QSOs, 3 normal galaxies, 8 starburst galaxies, 6
Galactic stars, and 2 candidates that are ``just'' visible on our
image.  The single candidates in these X-ray error circles have an
asterisk by their SOCA classification in Table\ref{table2}.

\section{Supplemental Data and Redshift Distributions}

\subsection{Radio and Near-IR identifications}

We have cross-correlated the positions of all of the X-ray sources
in our field-of-view with positions in the FIRST (Faint Images of
Radio Sources at Twenty centimeters) radio survey, the infrared
2MASS (2 Micron All Sky Survey) catalog, and the NED catalog. We
find seven of the X-ray error circles have FIRST radio sources
within them that are coincident with optical sources, in addition
to one associated with another radio source (HS0954+4815 as given
by NED; see Table~\ref{table2}). Of these 8 associated radio
sources, just two do not have SEDs in our data, and only three
have another object in their X-ray error circle.

There are a total of 100 radio sources in our image.  The
probability of having one radio source within 2$''$ of any optical
source in this image is 0.47 (taking the joint probability). This
makes the probability of having all eight position matches for the
radio sources with the optical sources at $2.3\times10^{-6}$ .
Hence, most of the radio sources are likely associated with their
optical counterparts.

In contrast, of the seven 2MASS object with position coincident
with our optical candidates, only 2 are classified as galaxies
[one normal galaxy, RXJ0953.8+4740(a) and one starburst galaxy,
RXJ0956.9+4731(a)], and five are identified with late-type stars
that are the dominant stellar candidates within these particular
X-ray error circles.

\subsection{Spectroscopic Observations and Objects of Special Note}

\begin{figure}
\plotone{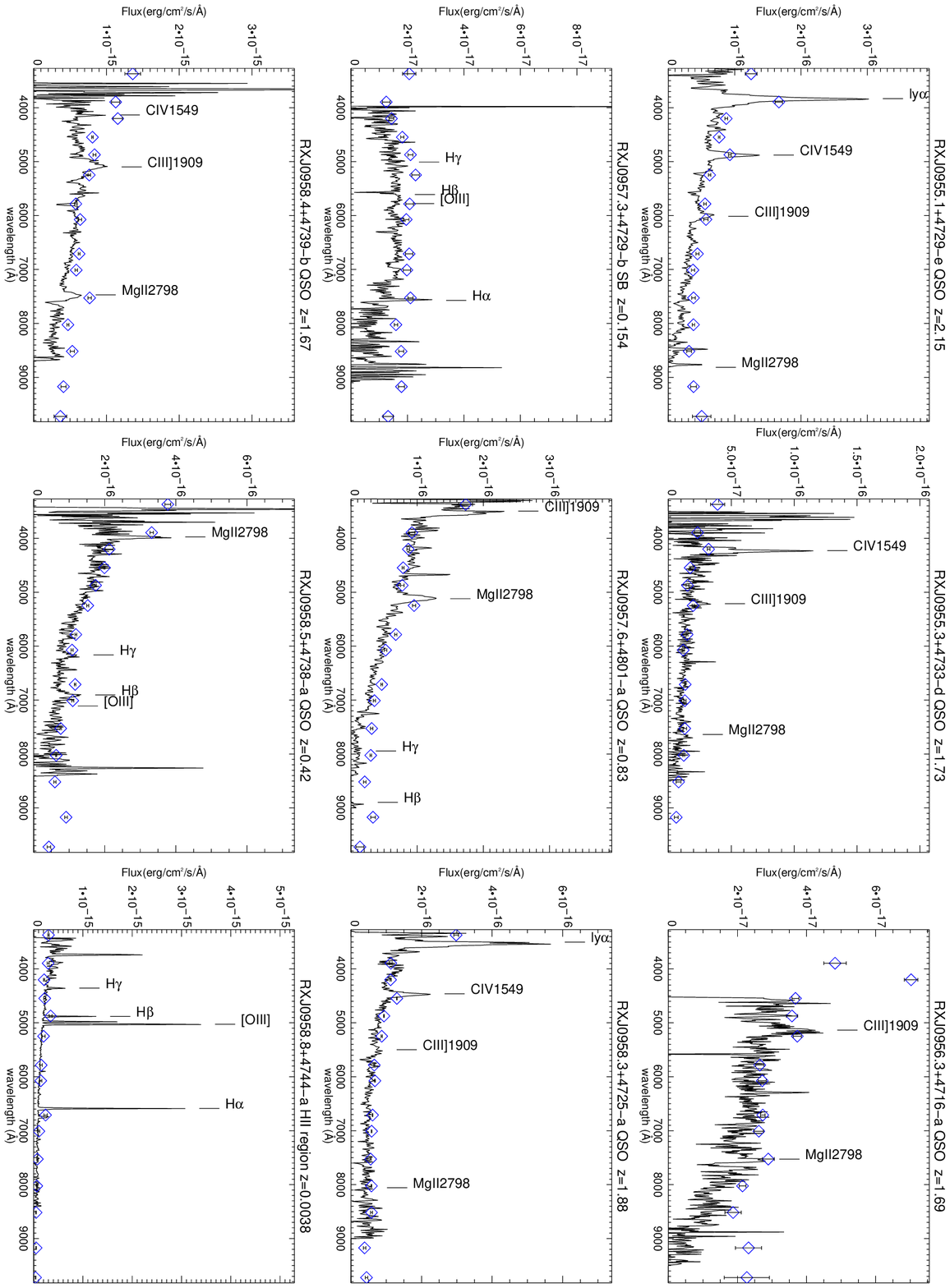} \caption{Spectra and photometric SEDs (points
with error bars) of some identified X-ray sources. It is shown
that photometric SEDs reflect well the main signatures of QSO and
starburst galaxy spectra.} \label{spec}
\end{figure}

We have obtained spectra of a subsample of our optically-detected
candidates to test our classification method and SOCA redshift
estimates. These spectra were taken with the slit spectrograph on
NAOC's 2.16 m telescope at its Xinglong Observing Station, and the
Multiobject Fiber Spectrograph (MOFS) on the 6m telescope of the
Special Astrophysical Observatory of the Russian Academy of
Sciences. Nine of these spectra are shown in Fig.\ref{spec}, with
the BATC fluxes overlayed. Included are 7 QSOs, 1 starburst galaxy
and the HII galaxy associated with UGC 5354. In addition, a search
of the SIMBAD and NASA Extragalactic Database (NED) catalogs comes
up with redshifts for two additional QSOs associated with these
X-ray sources (cf. Table~\ref{table2}).

The correspondence of spectroscopically-determined redshifts with
our SOCA-determined redshifts is excellent for 10 of the objects
(a mean error in z of 0.04 for these 10 objects), while it is off
by 1.03 for one additional object (RXJ0958.5+4738(a); shown in
Fig.~\ref{spec}). Examination in detail of the SOCA fitting
procedure for this one object shows that the BATC filter system
mistakes the emission line in the 9190{\AA} filter for [OII] 3727,
while the spectrum shows this must be H$\alpha$.

We have also made careful visual inspections for each of optical
counterparts within the X-error circles, relating their visual
appearance to the SEDs we obtain for them (something the reader
can also do, using Fig. \ref{sed} and Fig. \ref{idt}).  We give
special note to the objects found in the following X-ray sources:

RXJ0953.8+4740: There is evidently a galaxy group in this error
circle, comprised of objects b,c and d.  This group has previously
been identified as PDCS 36 (the 36th cluster/group found in the
Palomar Distant Cluster Survey, Postman et al. 1996).

RXJ0954.0+4756: This object is too faint to identify optically,
but its position is coincident with the radio source 7C0952+4814 =
FIRST J095401.1+475644.

RXJ0954.8+4715: The second object in Table~2 listed for this
source has a position coincident with the radio source CDS90-R307B
= FIRST J095453.2+471533.

RXJ0955.1+4729: There are 7 objects found in the X-ray error
circle. RXJ0955.1+4729(e) is a confirmed QSO at a redshift 2.15
and (d) is a late type star. What is interesting is that a,b,c,f
and g are classified as normal galaxies at redshifts
0.33,0.36,0.36,0.39 and 0.33 respectively. Visual inspection also
tends to put them at similar distances, making them a possible
galaxy group.

RXJ0956.7+4729: There are 3 objects within the error circle. From
the image we can see that all of them are within the diffraction
spikes of the nearby bright star, thus their SED are suspect.  As
a result, we put question marks in their identification in
Table~\ref{table2}.

RXJ0958.8+4744: This is part of the nearby, interacting galaxy UGC
5354.  Part of this galaxy system is a small, HII galaxy off to
one side.

RXJ09058.9+4745: This is a bright, cataloged F8 star, BD+48-1823.

\subsection{Redshift Distribution}

\begin{figure}
\epsscale{0.8}
\plotone{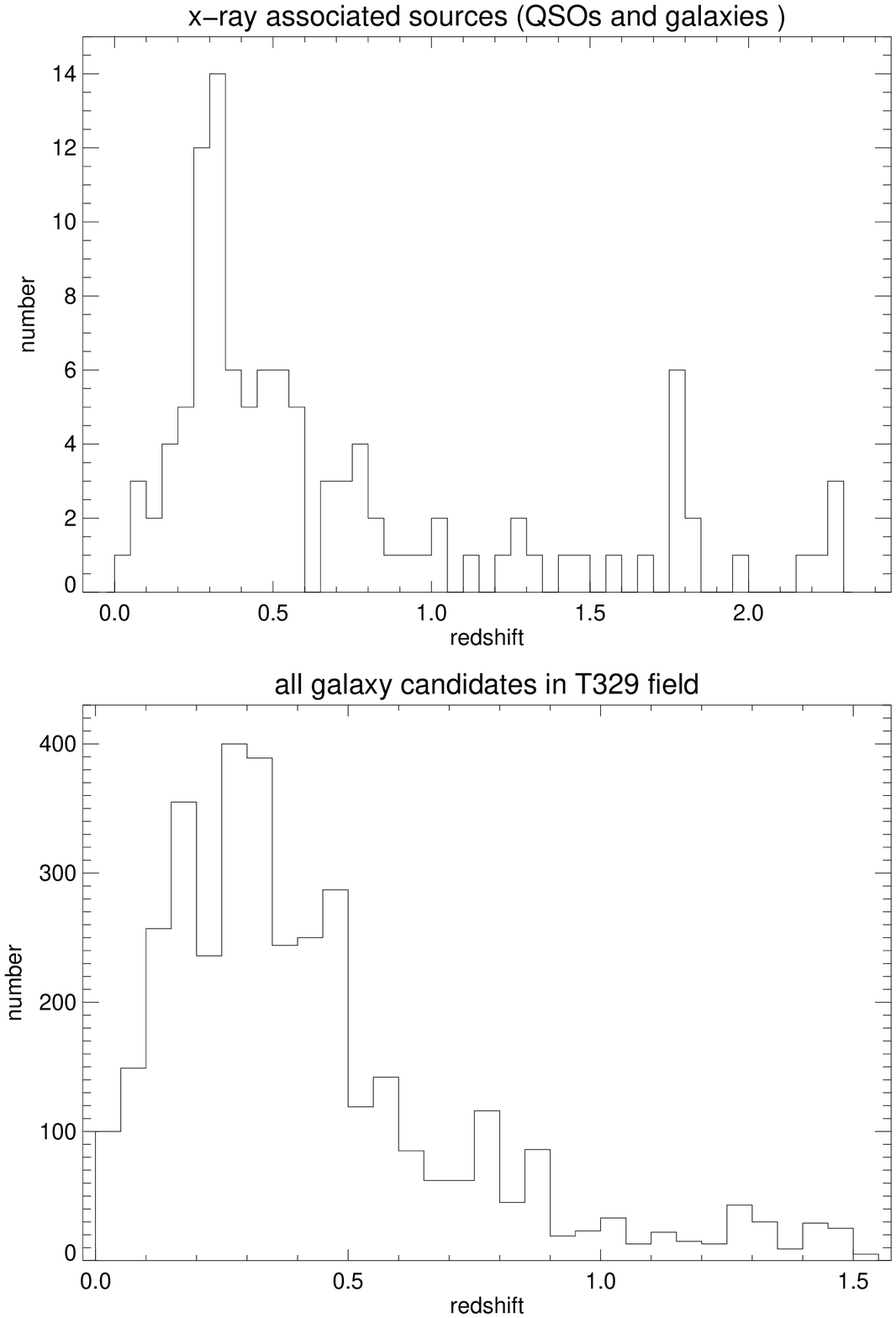}
%\plotone{zmd.ps}
\caption{Redshift distribution of the X-ray associated optical
sources (upper panel) and all the galaxy candidates in T329 field
(lower panel). The peak at 0.25$\sim$0.35 indicates several
superclusters along our line of sight at this redshift interval in
this field.} \label{redshift}
\end{figure}

In assembling the estimated redshifts for our X-ray associated
objects, we noticed that many of them tended to be clustered
around a redshift of 0.30$\pm 0.05$).  Given the accuracy of our
redshifts for galaxies ($<$0.1 for most individual objects), this
is significant. The top histogram in Fig~\ref{redshift} shows the
redshift distribution for the 110 galaxies plus QSOs X-ray source
candidates in our 58 arcmin$^2$ field of view.  It is evident that
there is a high overabundance of objects, mostly galaxies, in the
redshift range 0.25-0.35.  The bottom histogram in
Fig~\ref{redshift} shows the redshift distribution for all 3663
SOCA-classified galaxies in our field of view.

While the peak at redshift 0.25-0.35 is still there in the full
galaxy sample, the contrast of that peak appears to be more
significant for those galaxy candidates found in or near the error
circles for these X-ray sources. At this redshift, a degree-sized
field of view corresponds to a $\sim 20$ Mpc region of space. This
means that there is a collection of superclusters at this redshift
interval.  This is similar to what one would see if one looked
back at the local universe via sighting down the angle through the
Perseus cluster, the Local Supercluster, the Great Attractor and
the Shapley concentration stretched out over a redshift range of
nearly 20,000 km/sec.  In other words, seeing an overdense region
of galaxies over a redshift range of 0.1 is not that unusual in
our universe.

\section{Summary}

Based on the 15 color photometric observations, and SED-based
object classification approach (SOCA), as well as the
multiwavelength cross-correlations, we find 156 optical
candidates, to a magnitude limit close to $V \approx 23$, within
or near the error circles for all 75 X-ray sources in our field of
view. Among them are 31 QSOs (nine of which have spectroscopic
confirmation), 37 starburst galaxies (2 of which have
spectroscopic confirmation), 42 normal galaxies (including 3
possible galaxy groups), 39 stars (one of which is a BD object),
and 7 just ``visible'' objects with no classifications (two of
which are coincident with known radio sources). Two of the X-ray
error circles have only just ``visible'' objects in them. We find
8 radio sources (out of 100 in our image) that are coincident with
an optical object within the X-ray error circles, making it likely
that many of these are the optical counterparts to these radio
sources.

Separately, we have also SED-classified 6011 additional objects in
our 3364 arcmin$^2$ field of view to the same apparent magnitude
limit (i.e., not including those found near or in the X-ray error
circles). Of these 6011 objects, 341 are QSOs, 1912 are stars,
1508 are starburst galaxies, and 2076 are galaxies and 174 are
unclassified for a number of objective reasons.  The area of our
X-ray circles subtends 31.52 arcmin$^2$, or 0.00937 of our full
3364 arcmin$^2$ field of view.  Considering the objects detected
to objects that could be randomly found in these error circles, we
have: $6.5\sigma$ detections for all 149 classified objects,
$4.7\sigma$ for the QSOs, and from $2.8-3.2\sigma$ for stars,
starburst galaxies and galaxies. Twenty-eight error circles have
only one object in or near them, including: 9 QSOs, 3 normal
galaxies, 8 starburst galaxies, 6 stars and 2 ``just'' visible
objects.

In sum, in this paper we perform an exercise that few have been
able to do with their X-ray data. By being able to SED-classify
all objects in our 3364 arcmin$^2$ field of view down to V $\sim$
23, we can ask ourselves what are the kinds of optical candidates
we find within X-ray error circles to those randomly found in the
field of view. question is that while all classified objects:
QSOs, starburst galaxies, normal galaxies and stars are
overrepresented in the X-ray error circles compared to a random
distribution, it is only the QSOs that are highly statistically
found in these X-ray circles.  Yet, at the same time, there are
6.5$\sigma$ more objects within these X-ray circles than if
randomly distributed in this image. So, while we know that one of
the objects in our X-ray error circles is likely the X-ray source,
in absence of independent knowledge, choosing which object it is
is still more arbitrary than scientific.

\acknowledgments

The BATC Survey is supported by the Chinese Academy of Sciences,
the Chinese National Natural Science Foundation and the Chinese
State Committee of Sciences and Technology. The present work was
partially supported by the Chinese National Key Basic Research
Science Foundation (NKBRSFG19990754).  This research has made use
of the NASA/IPAC Extra-galactic Database (NED) which is operated
by the Jet Propulsion Laboratory, California Institute of
Technology, under contract with the National Aeronautics and Space
Administration.

\input table1
\clearpage
\input table2

\clearpage
\clearpage
%\clearpage
\begin{figure}
\plotone{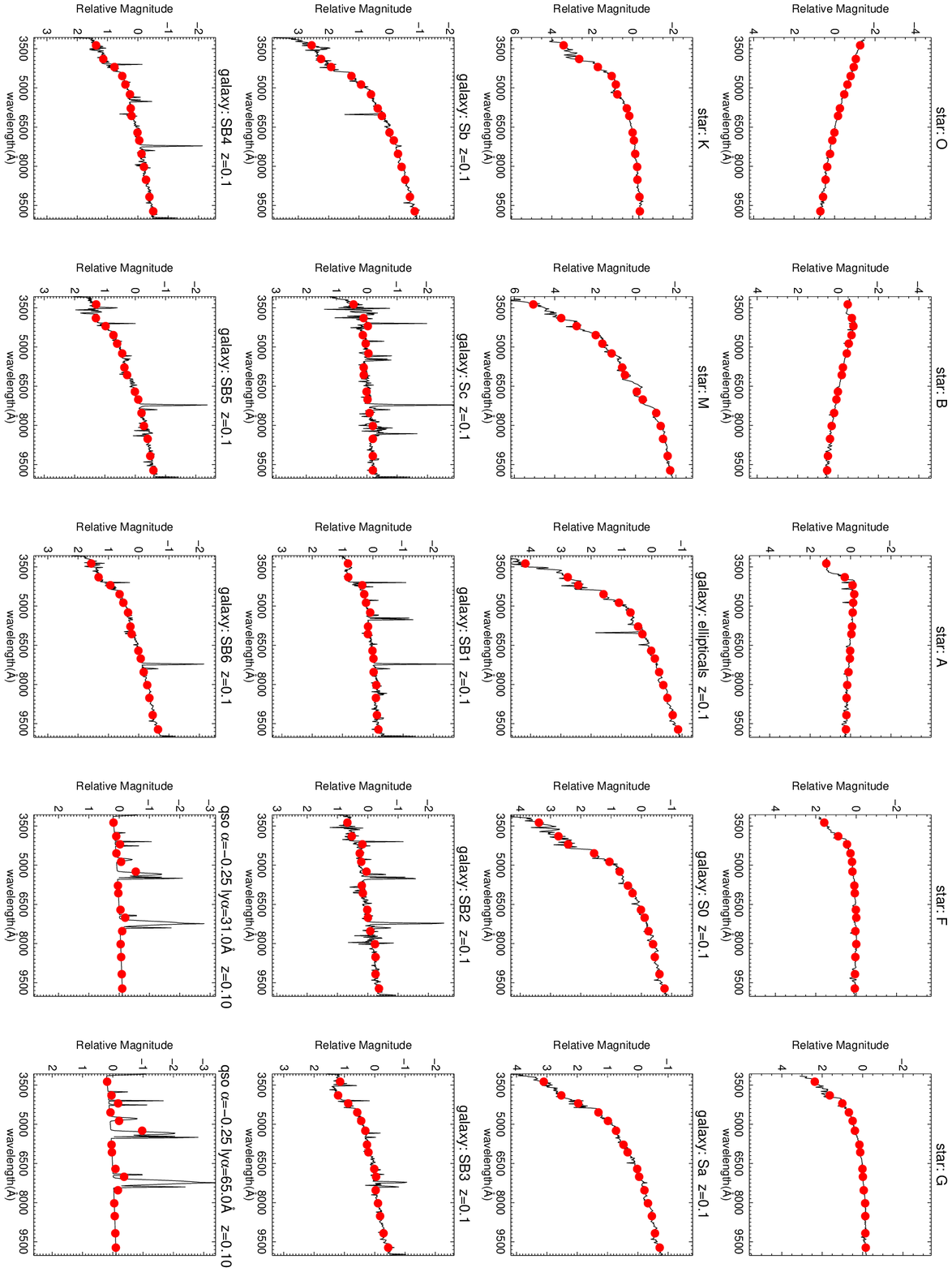} \caption{The template spectra as well as the
corresponding BATC SED (filled circle) used in the SOCA.
Classification type and redshift (for extragalactic objects, see
text) are as marked in the title. } \label{f2}
\end{figure}
\clearpage
\begin{figure}
%\epsscale{0.85}
\figurenum{3} \plotone{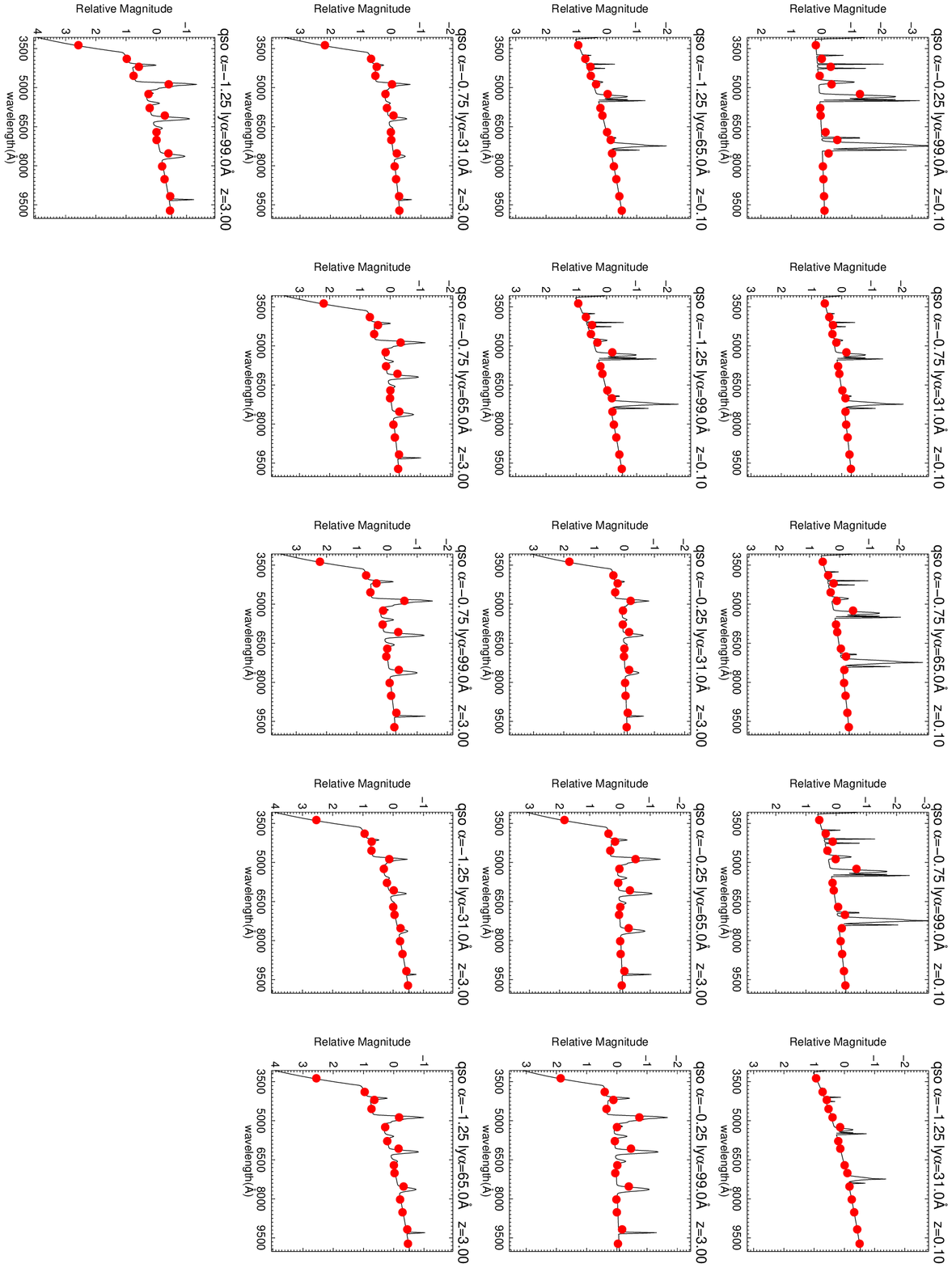} \caption{continued}
\end{figure}

\clearpage
\begin{figure}
\plotone{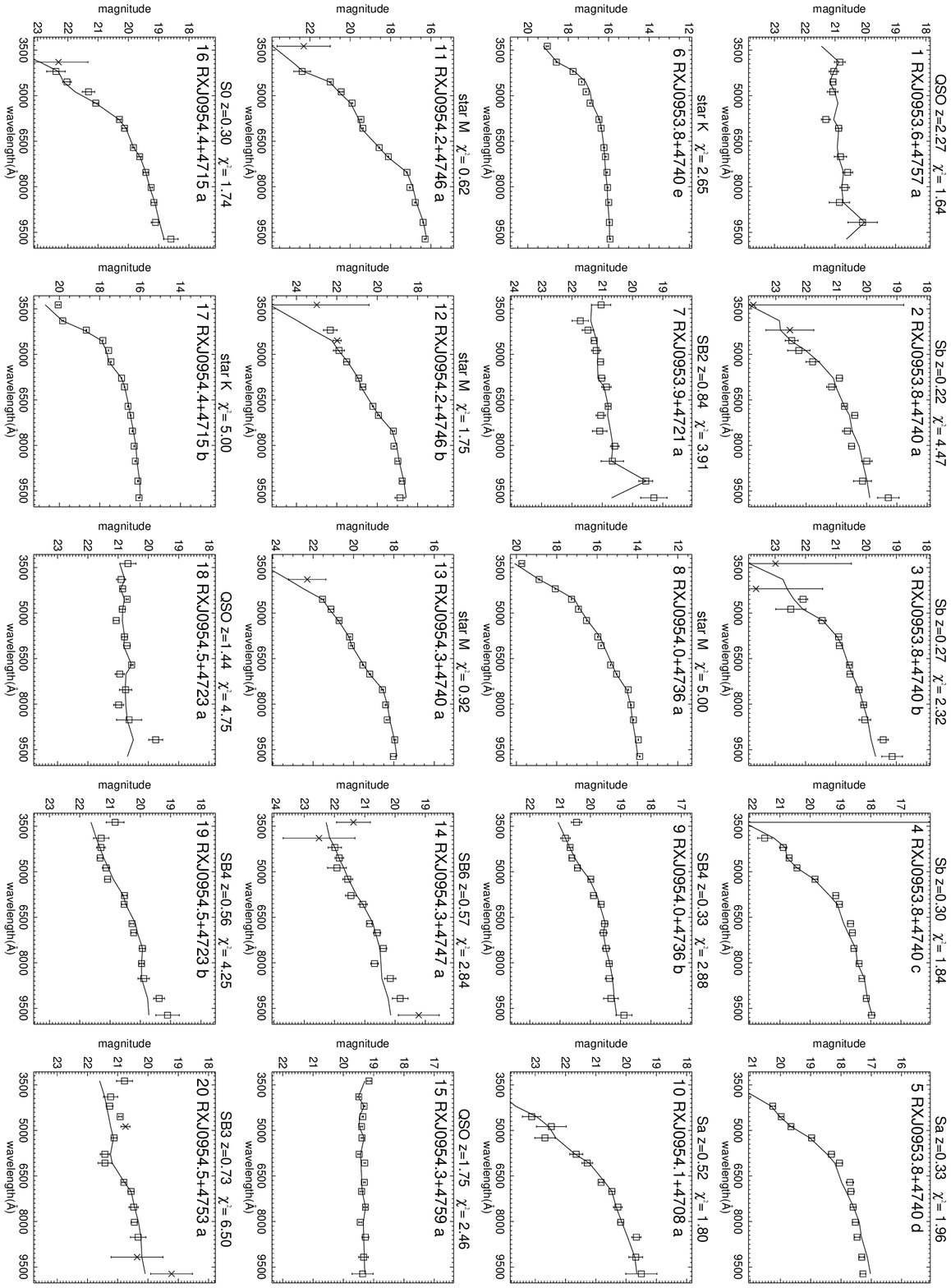} \caption{\tiny{ SED plots as well as the best
template fits are shown for all the spatially-associated optical
sources for 73 of the 75 X-ray sources. Solid lines: best fitted
template SED; squares: BATC PSF photometry SED; crosses: aperture
photometry result for the filters is too faint to perform accurate
photometry; error bars: photometry errors (see Zhou et al. 2002
for detail).  Note that we do not use data that have great
uncertainty in the SOCA classification. The best fitted $\chi^2$
is also listed in for each SED.}} \label{sed}
\end{figure}

\clearpage
\begin{figure}
\figurenum{4} \plotone{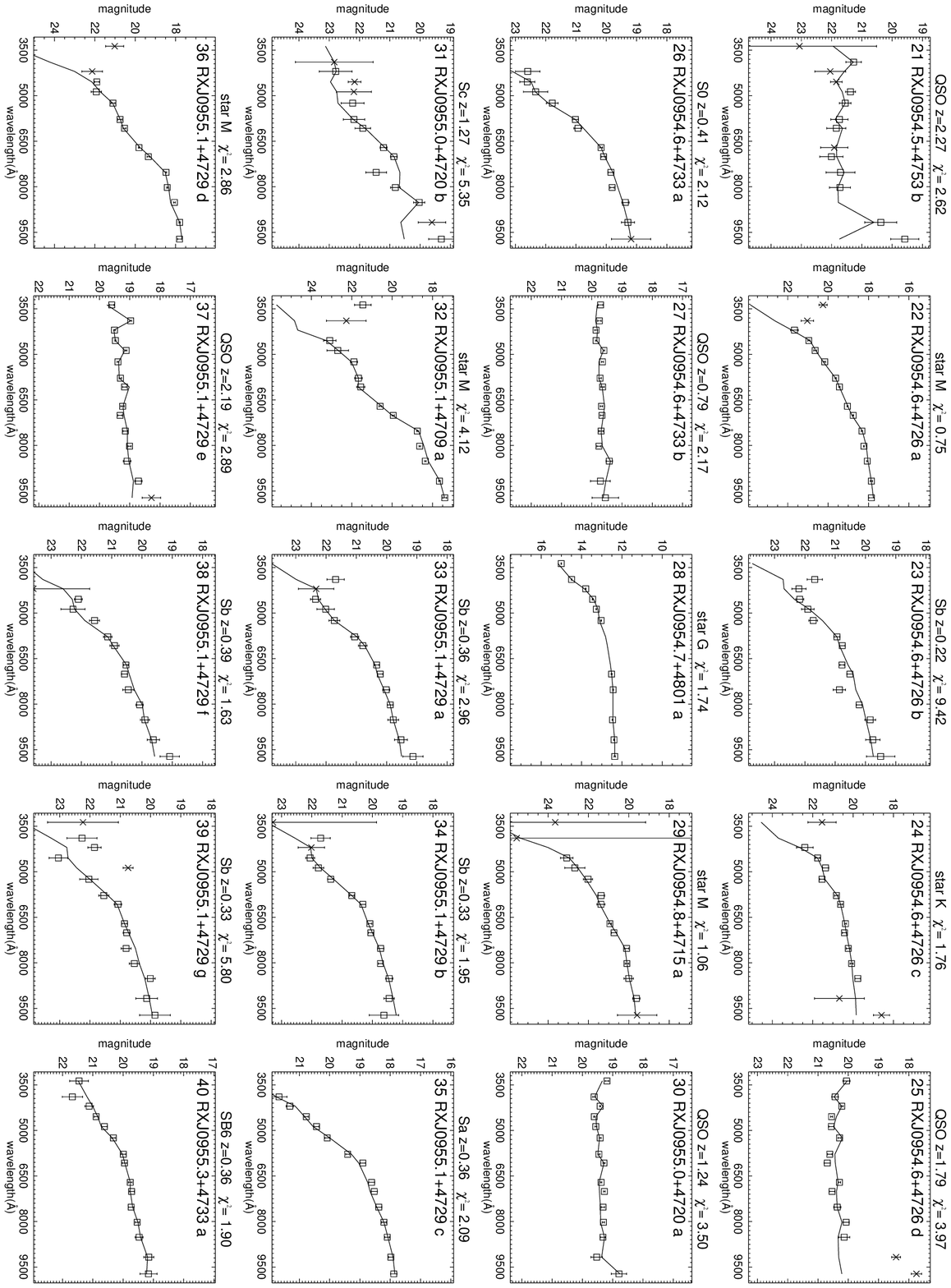} \caption{continued}
\end{figure}
\clearpage
\begin{figure}
\plotone{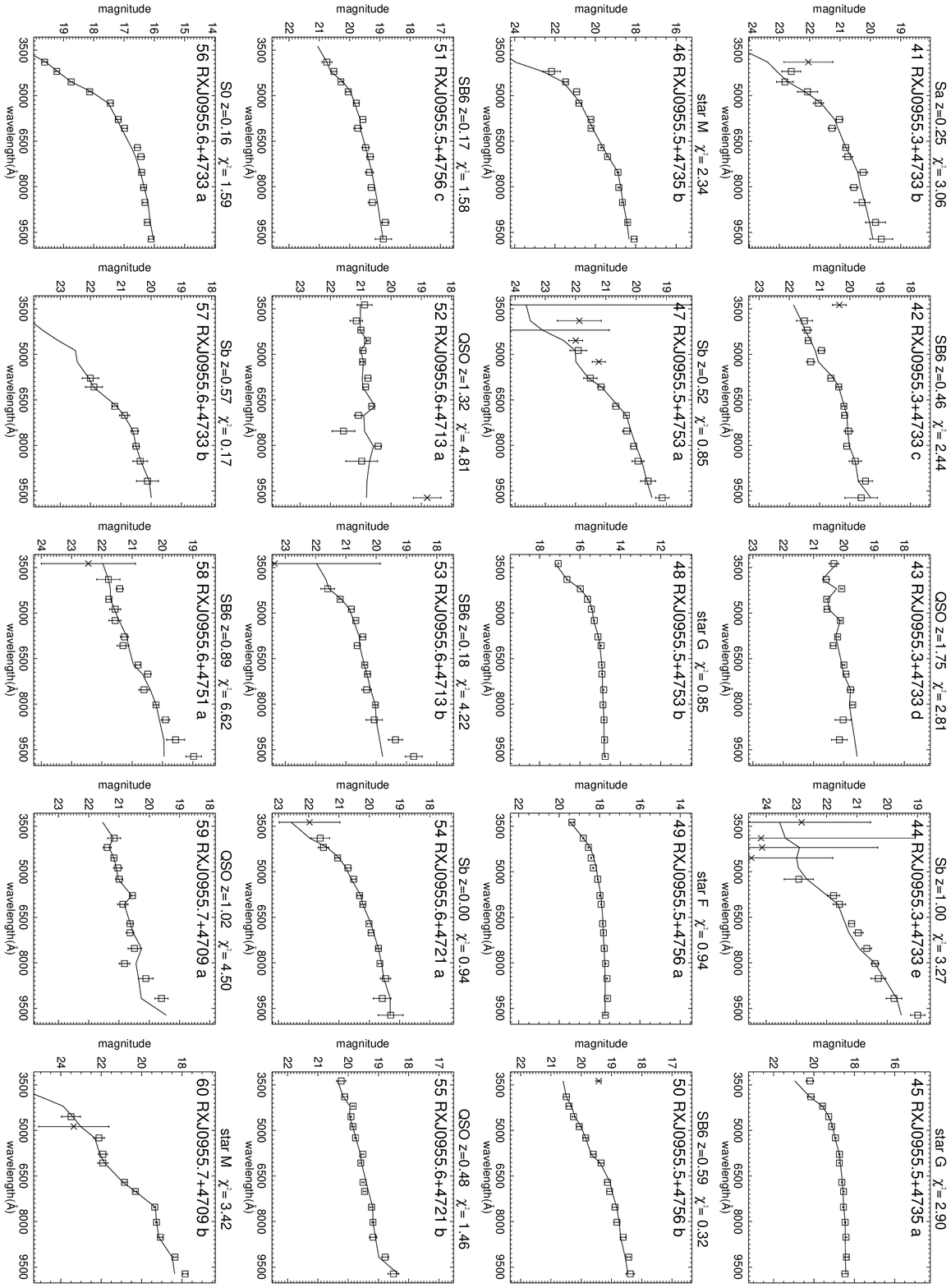} \figurenum{4} \caption{continued}
\end{figure}
\clearpage
\begin{figure}
\plotone{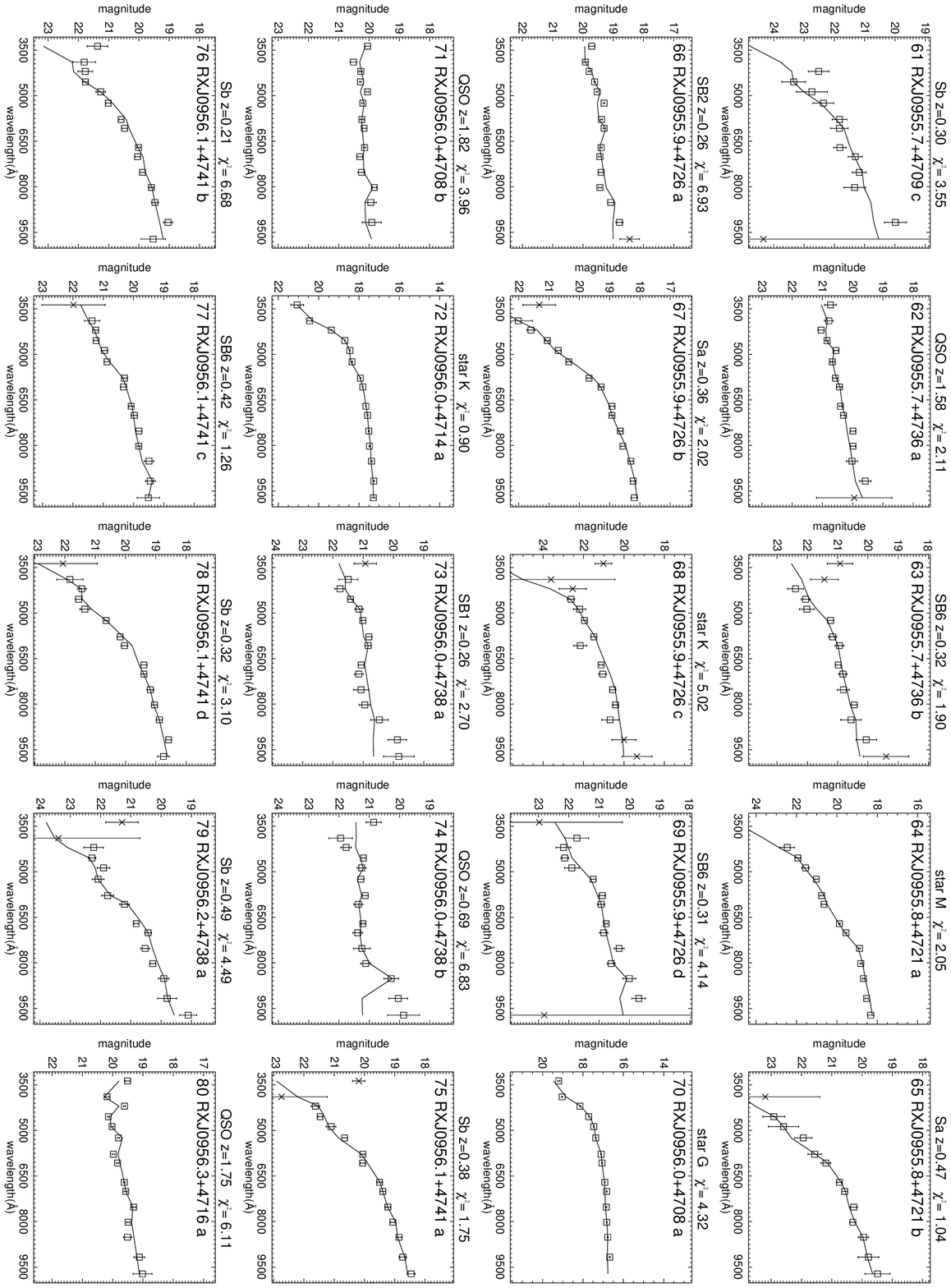} \figurenum{4} \caption{continued}
\end{figure}
\clearpage
\begin{figure}
\plotone{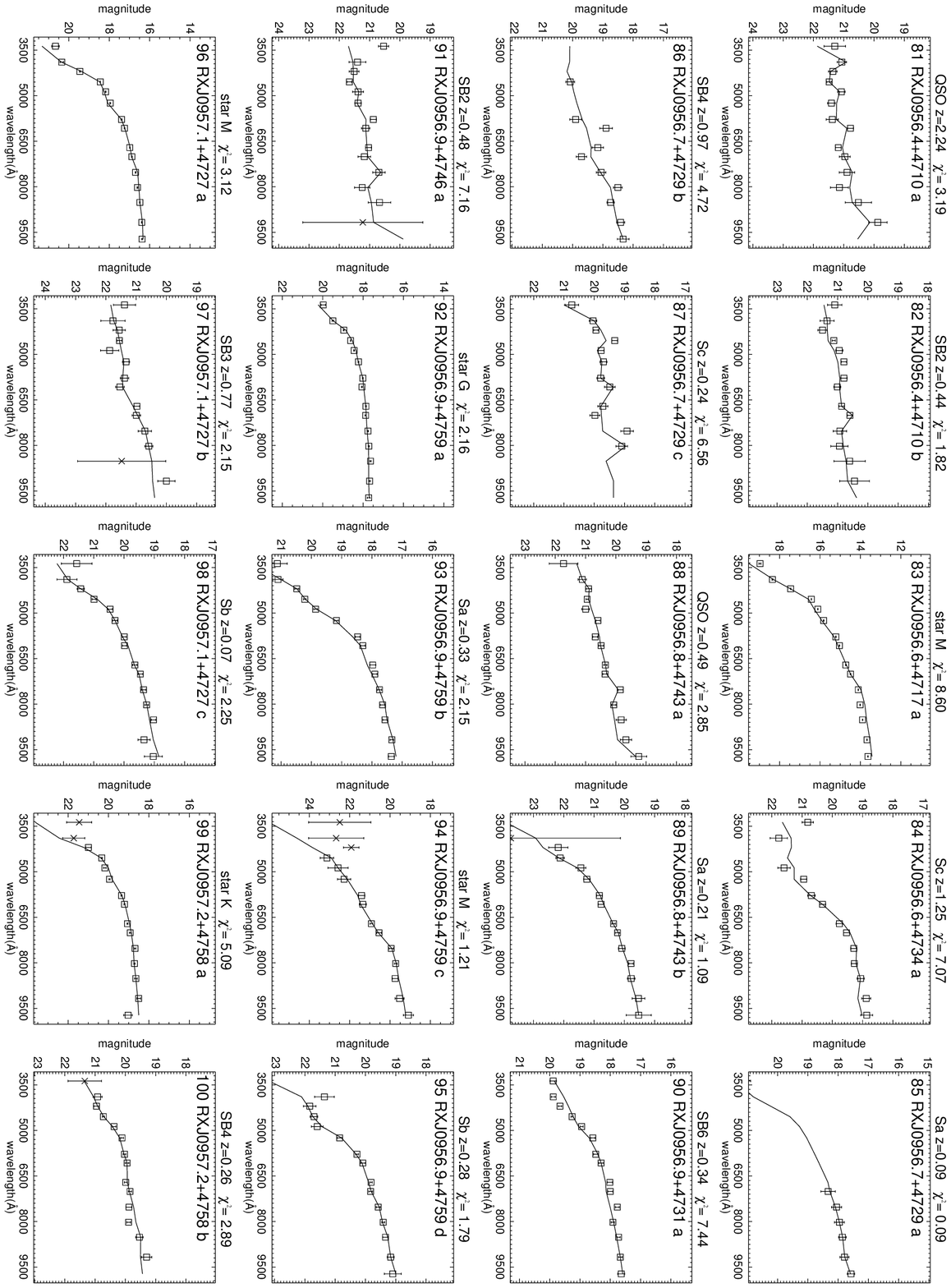} \figurenum{4} \caption{continued}
\end{figure}
\clearpage
\begin{figure}
\plotone{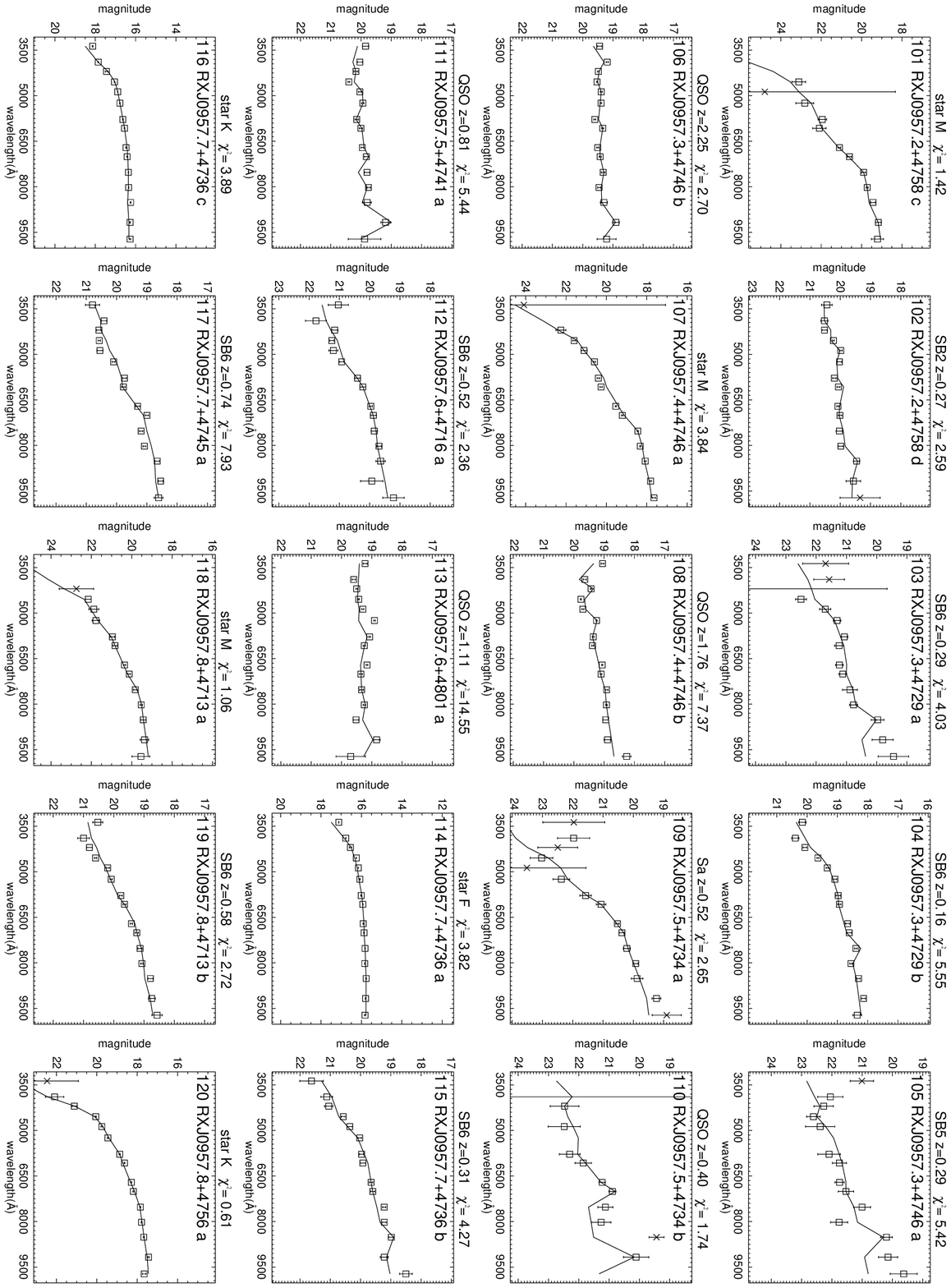} \figurenum{4} \caption{continued}
\end{figure}
\begin{figure}
\plotone{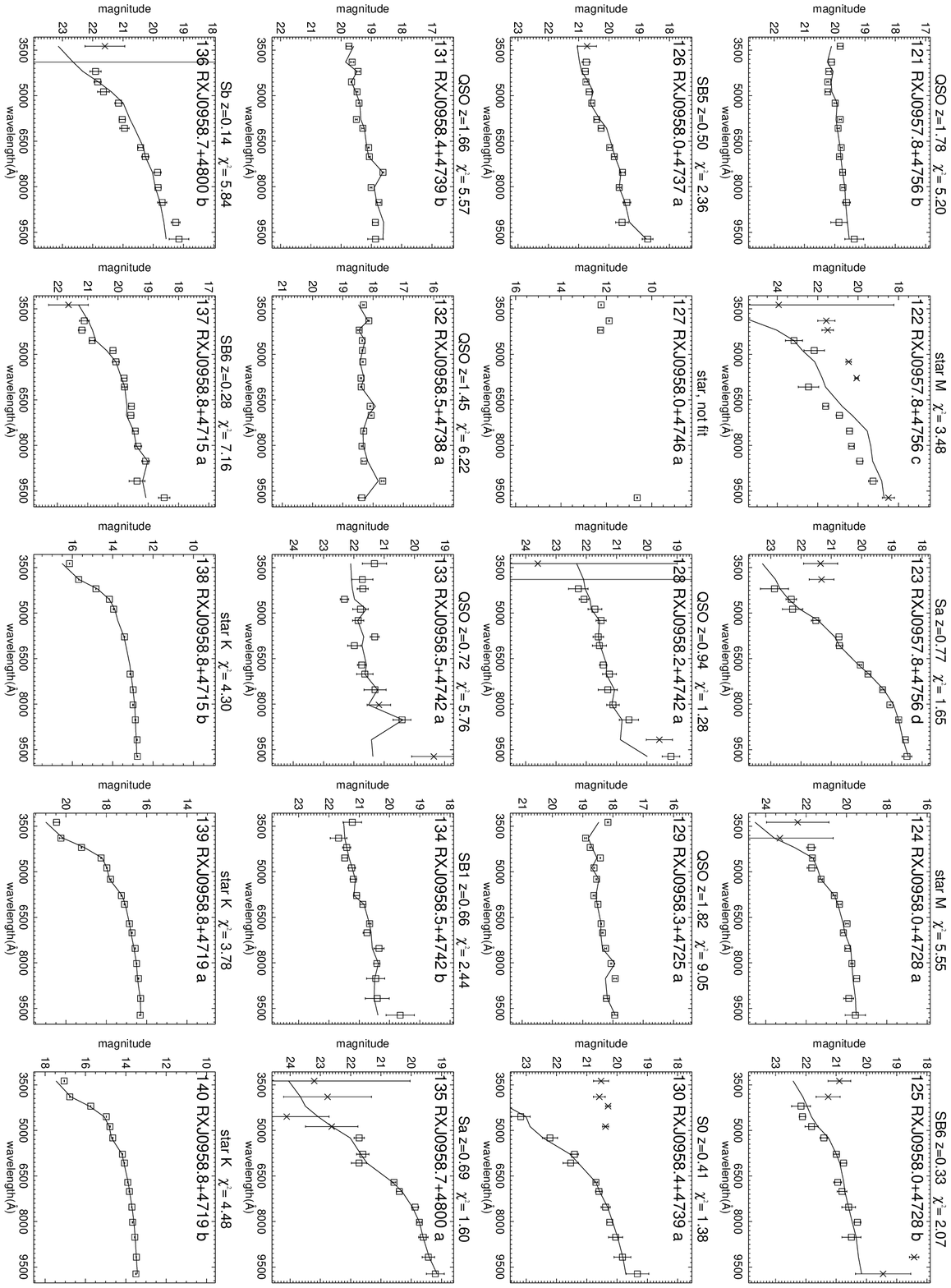} \figurenum{4} \caption{continued}
\end{figure}
\clearpage
\begin{figure}
\plotone{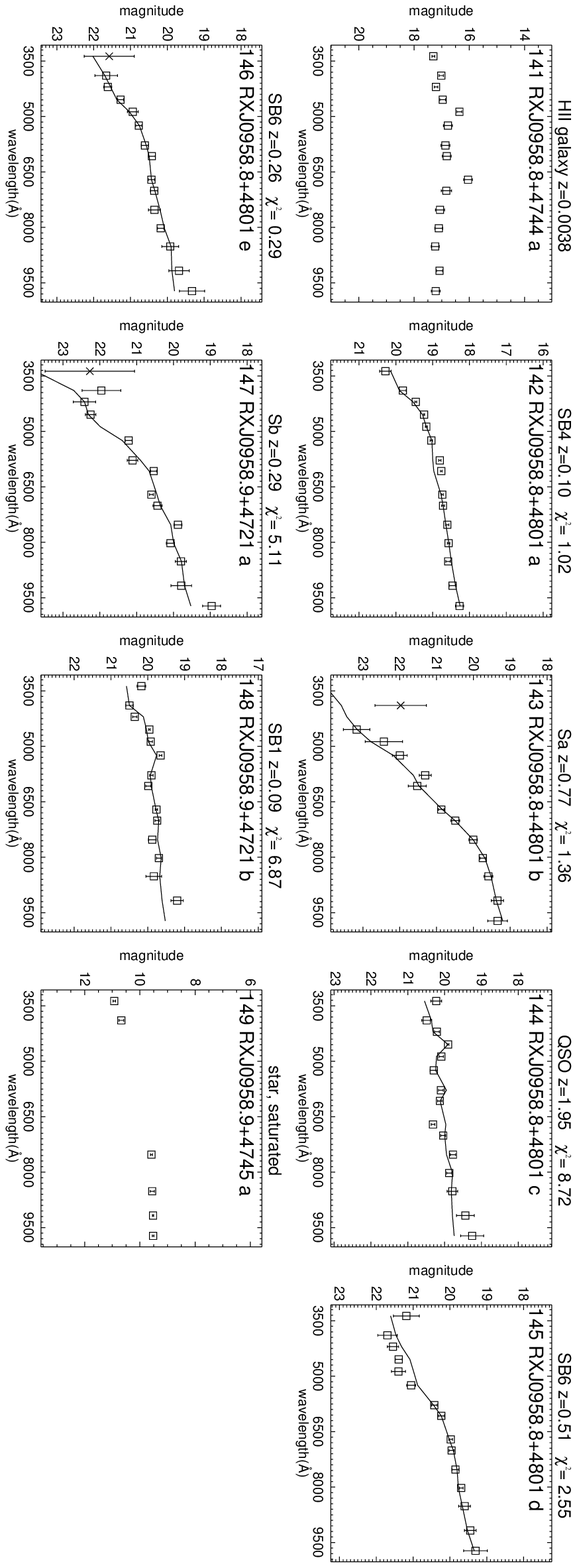} \figurenum{4} \caption{continued}
%\epsscale{0.85}

\end{figure}

\clearpage
\begin{figure}
\epsscale{0.8}
\plotone{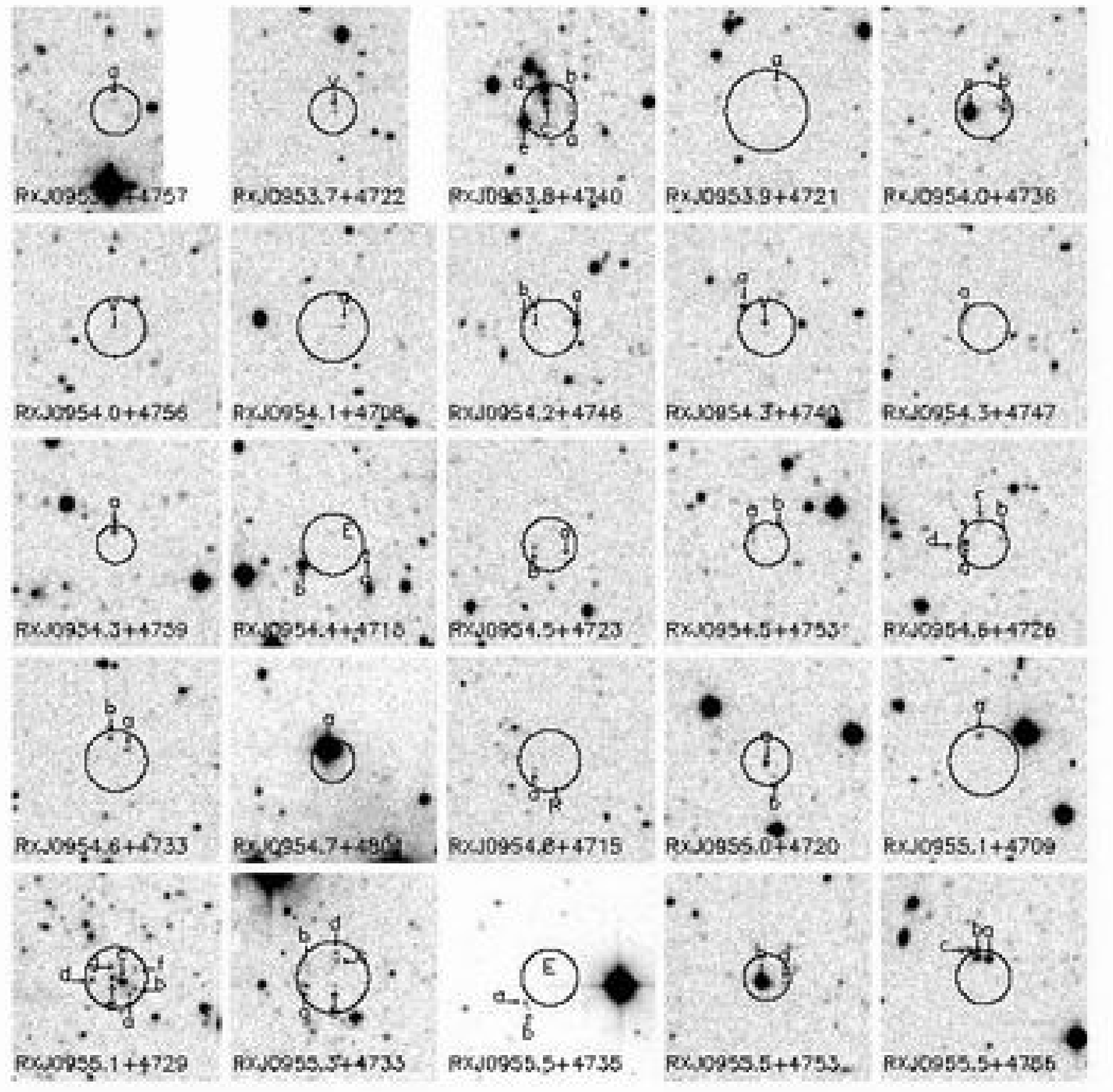} \caption{ J-band finding charts of all the 75
X-ray sources in the 1 deg$^2$ field. Each chart is in size of
2.8'$\times$2.8', north is up and east to the left. The different
size of circles indicates X-ray detection error in radius of
$R=2\sigma+10"$ (Multhagen et al. 1997). E denotes Empty in the
error circle, V denotes a visible only source, a-h are individual
objects whose data are given in table~\ref{table2}.} \label{idt}
\end{figure}
\clearpage
\begin{figure}
\plotone{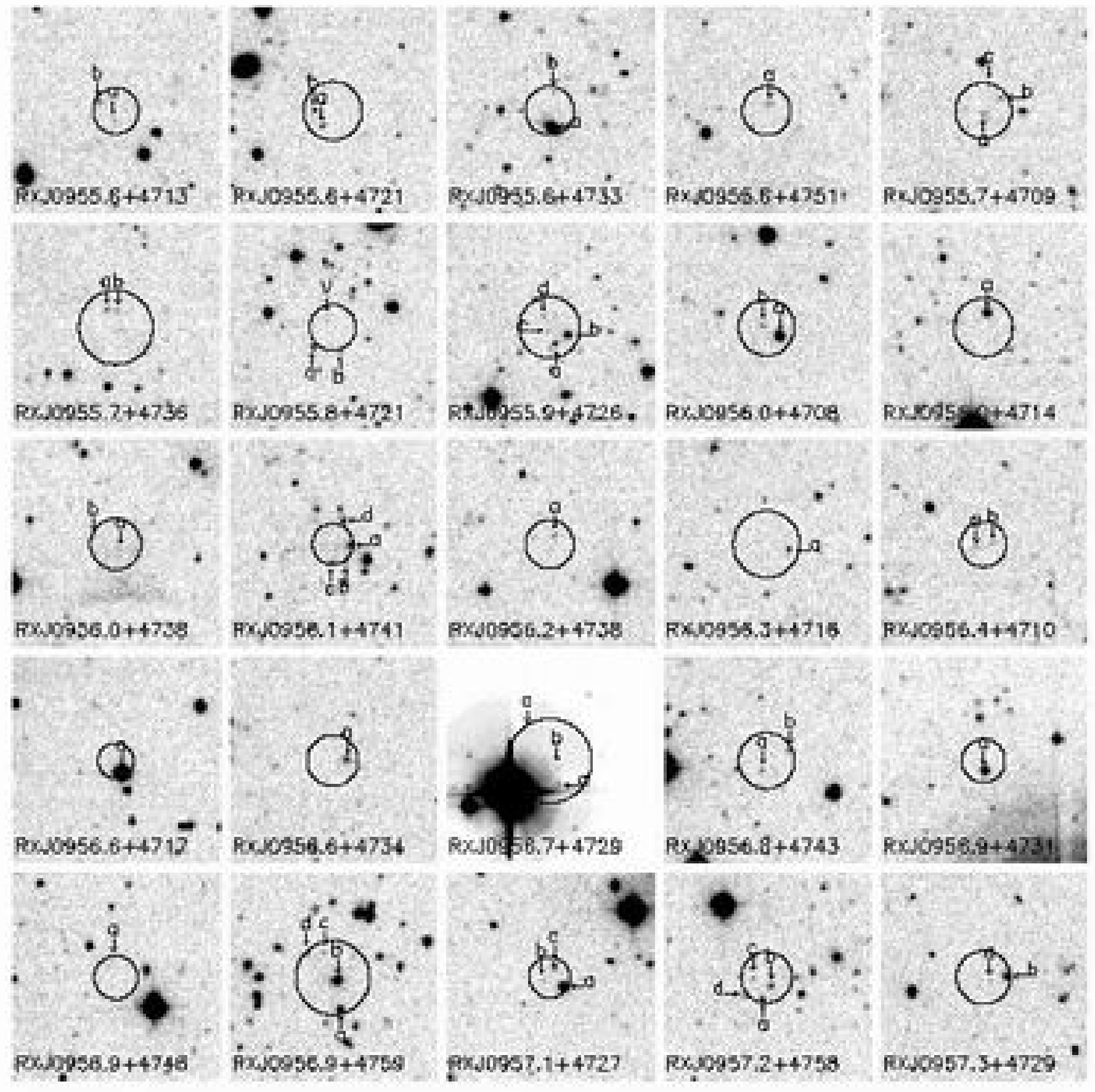} \figurenum{5} \caption{continued}
\end{figure}
\clearpage
\begin{figure}
\plotone{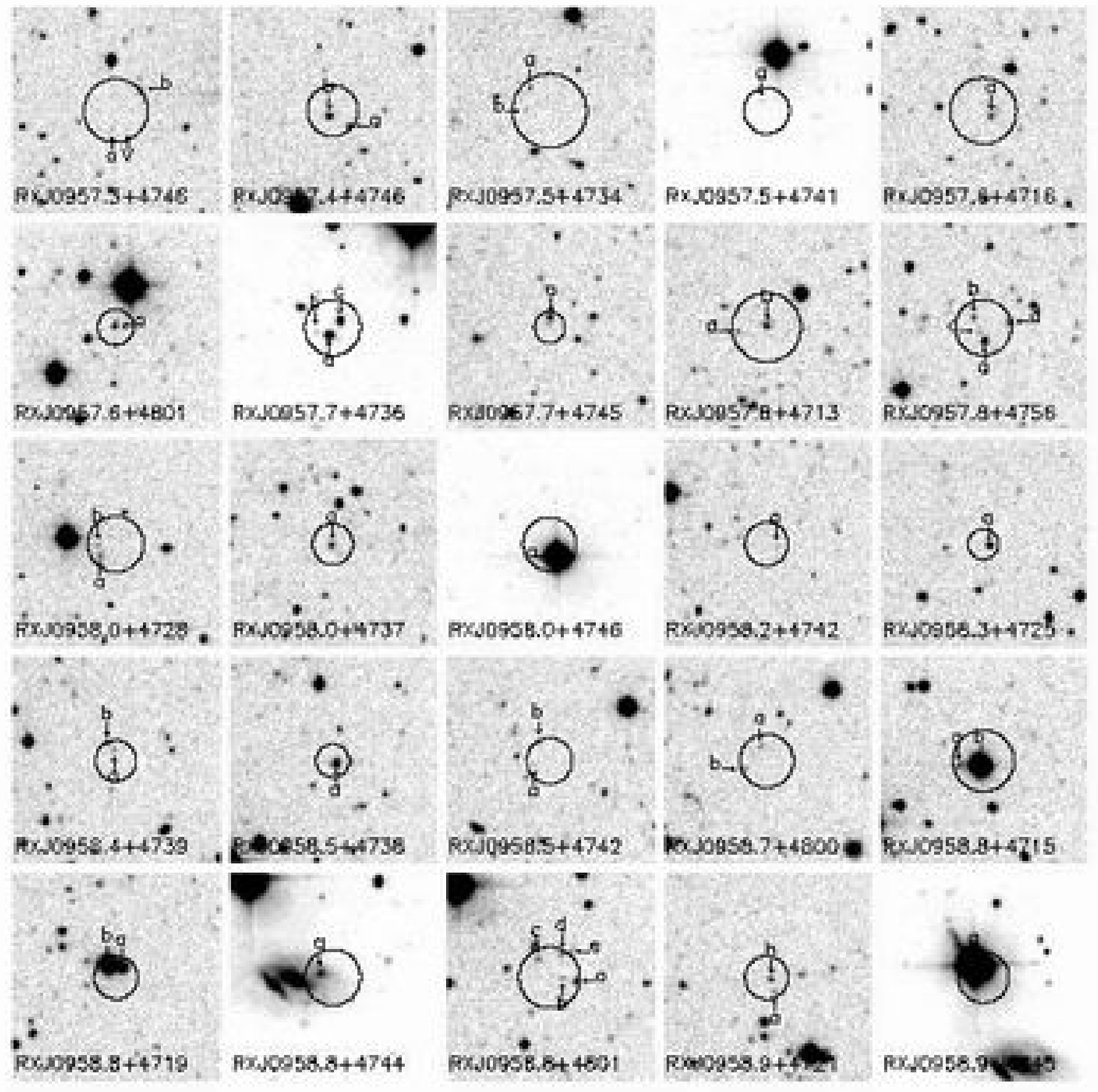} \figurenum{5} \caption{continued}
\end{figure}

\clearpage
\clearpage
\clearpage
\end{document}

%% file: table1.tex
\begin{deluxetable}{cccccc}
%\tablewidth{0pt}
%\small
\tiny
%\tablenum{1}
\tablecaption{Optical Observation Log \label{table1}} \tablehead{
Filter &Wavelength~$\lambda$ &Frames & Exposure & Calibrations &
Limiting Magnitude  } \startdata a & 3360 & 7 & 03:04 & 1 & 22.13
\nl b & 3890 & 12 & 03:44 & 1 & 22.56  \nl c & 4210 & 19 & 08:45 &
7 & 23.23  \nl d & 4550 & 35 & 09:58 & 5 & 23.68  \nl e & 4920 &
29 & 03:04 & 10& 22.83  \nl f & 5270 & 25 & 06:00 & 5 & 23.22  \nl
g & 5795 & 30 & 06:54 & 5 & 22.83  \nl h & 6075 & 21 & 04:44 & 3 &
22.85  \nl i & 6660 & 31 & 05:49 & 10& 22.61  \nl j & 7050 & 27 &
06:53 & 5 & 22.55  \nl k & 7490 & 22 & 05:52 & 22 &21.91  \nl m &
8020 & 22 & 06:50 & 6 & 22.07  \nl n & 8480 & 12 & 04:50 & 3 &
21.34  \nl o & 9190 & 13 & 04:41 & 1 & 20.81  \nl p & 9745 & 12 &
05:00 & 2 & 20.24  \nl
%\tablenotetext{e}{Object classifications based on the SED information.}
%\tablecomments{}
\enddata
\end{deluxetable}

%% file: table2.tex
\begin{deluxetable}{lrrllllrrrll}
\tablewidth{0pt}
%\small
\tabletypesize{\tiny}
%\tablenum{1}
\tablecaption{X-ray sources and Optical Identifications in 1
deg$^2$ Field \label{table2}} \tablehead{ ROSAT source
&Cts&X\tablenotemark{a}$_{err}$ & Old ID\tablenotemark{b} &
\multicolumn{3}{c}{Associated Optical Objects}
 & $\Delta r$\tablenotemark{d} &m$_V$\tablenotemark{e}&$f_{\rm xo}$\tablenotemark{f} &
Pred z\tablenotemark{g} & New ID\tablenotemark{h} \nl \cline{5-7}
&$10^{-2}$ & $\arcsec$  & &obj\tablenotemark{c} & RA(J2000) &
DEC(J2000) &  $\arcsec$ & & & }

\startdata RXJ0953.6+4757&3.9&19&D      &a&09:53:37.94&47:57:33.6&
7.60&22.12   & 1.63   &2.27     &*QSO   \nl
RXJ0953.7+4722&5.7&18&D      &V&09:53:47.20&47:21:57.7&
2.51&$>$23.01&$>$2.15&$\ldots$  &*?     \nl
RXJ0953.8+4740&6.9&21&D      &a&09:53:51.92&47:40:00.7&18.50&21.24
& 1.52   &0.22     &Sb     \nl
             &&  &           &b&09:53:51.96&47:40:07.5&17.80&21.14   & 1.48   &0.27     &Sb     \nl
             &&  &           &c&09:53:53.94&47:40:10.3& 5.80&19.43   & 0.80   &0.30     &Sb     \nl
             &&  &           &d&09:53:54.04&47:40:22.5&17.80&18.59   & 0.46   &0.33     &Sa    \\ % \nl
             &&&&&&&&&& \multicolumn{2}{c}{b,c,d grp=PDCS 36} \\ % \nl
             &&  &           &e&09:53:55.78&47:39:54.9&23.31&16.71   &-0.29  &$\ldots$  &K star \nl
RXJ0953.9+4721&11.3&32&unid  &a&09:53:53.72&47:22:15.8&19.00&21.09
& 1.67  &0.84      &*SB2    \nl RXJ0954.0+4736&2.9&22&M can
&a&09:54:01.90&47:36:03.5&11.40&16.20   &-0.86  &$\ldots$  &M star
\nl
             &&  &           &b&09:53:59.09&47:36:04.4&17.40&19.98   & 0.64  &0.33      &SB4    \nl
RXJ0954.0+4756&3.6&23&D    &V,R&09:54:01.18&47:56:44.1&
7.17&$>$23.01&$>$1.95&$\ldots$  &*?  \nl
             &&&&&&&\multicolumn{4}{c}{7C0952+4814, FIRST J095401.1+475644} \\
RXJ0954.1+4708&2.6&27&AGN can&a&09:54:10.99&47:08:45.0& 9.30&22.04
& 1.42  &0.52      &*Sa     \nl RXJ0954.2+4746&2.7&22&empty
&V&09:54:14.99&47:46:45.1&12.96&$>$23.01&$>$1.95&$\ldots$  &?
\nl
             &&  &           &a&09:54:11.57&47:46:51.6&22.53&19.70   & 0.50  &$\ldots$  &M star\nl
             &&  &           &b&09:54:15.87&47:46:56.3&21.79&21.23   & 1.11  &$\ldots$  &M star \nl
RXJ0954.3+4740&1.8&22&D      &V&09:54:21.95&47:40:52.0&
5.22&$>$23.01&$>$1.65&$\ldots$  &?       \nl
             &&  &           &a&09:54:23.61&47:41:14.1&25.03&20.45   & 0.63  &$\ldots$  &M star\nl
RXJ0954.3+4747&4.3&19&D      &a&09:54:24.96&47:47:07.6&15.40&21.55
& 1.44  &0.57      &*SB6    \nl RXJ0954.3+4759&3.7&15&AGN
can&a&09:54:20.26&47:59:43.8& 9.80&19.51   & 0.56  &1.75
&*QSO    \nl RXJ0954.4+4715&5.6&24&unid.
&a&09:54:25.57&47:15:34.3&28.48&20.66   & 1.20  &0.30      &S0
\nl
             &&  &           &b&09:54:30.74&47:15:26.8&31.06&17.20   &-0.18  &$\ldots$  &K star\nl
RXJ0954.5+4723&7.1&21&D      &a&09:54:30.56&47:23:02.4&20.00&20.94
& 1.41  &1.44      &QSO    \nl
             &&  &           &b&09:54:33.08&47:23:14.4&13.50&20.76   & 1.34  &0.56      &SB4    \nl
RXJ0954.5+4753&2.5&17&D      &a&09:54:31.53&47:53:40.4&11.60&21.38
& 1.13  &0.73      &SB3    \nl
             &&  &           &b&09:54:29.37&47:53:45.9&13.03&21.71   & 1.27  &2.27      &QSO    \nl
RXJ0954.6+4726&8.5&19&D      &a&09:54:41.75&47:26:20.7&14.70&19.86
& 1.06  &$\ldots$  &M star\nl
             &&  &           &b&09:54:38.76&47:26:23.6&16.00&21.25   & 1.61  &0.22      &Sb     \nl
             &&  &           &c&09:54:40.51&47:26:36.1&16.20&21.12   & 1.56  &$\ldots$  &K star \nl
             &&  &           &d&09:54:42.38&47:26:17.6&21.24&20.54   & 1.34  &1.79      &QSO    \nl
RXJ0954.6+4733&2.8&24&AGN can&a&09:54:39.11&47:33:54.2&12.30&21.33
& 1.16  &0.41      &S0      \nl
             &&  &           &b&09:54:40.46&47:34:05.0&18.40&19.76   & 0.46  &0.79      &QSO    \nl
RXJ0954.7+4801&3.0&17&sat    &a&09:54:47.38&48:01:28.0&
9.40&$<$13.20&$<$-2.05&$\ldots$ &*G star \nl
RXJ0954.8+4715&2.9&24&D      &a&09:54:54.82&47:15:43.6&14.50&21.64
& 1.30  &$\ldots$  &M star  \nl
             &&  &           &V,R&09:54:53.17&47:15:33.1&15.20&$>$23.01&$>$1.86&$\ldots$&?      \nl
             &&&&&&&&\multicolumn{4}{c}{CDS90-R307B, FIRST J095453.2+471533}\nl
RXJ0955.0+4720&3.7&19&AGN    &a&09:55:02.19&47:20:08.6& 4.40&19.50
& 0.55  &1.24      &QSO, z=1.22\tablenotemark{j}  \nl
             &&  &           &b&09:55:01.61&47:19:57.5&16.62&22.35   & 1.70  &1.27      &Sc?     \nl
RXJ0955.1+4709&0.8&27&D      &a&09:55:06.46&47:10:00.2&20.40&21.80
& 0.81  &$\ldots$  &*M star? \nl RXJ0955.1+4729&2.4&24&AGN
&a&09:55:05.60&47:28:54.2&19.60&21.33   & 1.10  &0.36      &Sb
\nl
             &&  &           &b&09:55:05.12&47:28:59.1&19.10&20.96   & 0.95  &0.33      &Sb      \nl
             &&  &           &c&09:55:06.08&47:29:05.2& 7.80&19.68   & 0.44  &0.36      &Sa      \nl
             &&  &           &d&09:55:08.55&47:29:07.6&20.10&20.93   & 0.94  &$\ldots$  &M star  \nl
             &&  &           &e&09:55:06.91&47:29:08.8& 3.80&19.39   & 0.32  &2.19      &QSO, z=2.15\tablenotemark{j}  \nl
             &&  &           &f&09:55:05.04&47:29:15.3&16.40&21.33   & 1.10  &0.39      &Sb      \nl
             &&  &           &g&09:55:06.30&47:29:18.3& 7.91&21.91   & 1.34  &0.33      &Sb      \nl
             &&&&&&&&&&\multicolumn{2}{c}{a,b,c,f,g grp?}\nl
RXJ0955.3+4733&5.6&28&D      &a&09:55:21.65&47:33:47.9&24.20&20.16
& 1.00  &0.36      &SB6     \nl
             &&  &           &b&09:55:21.56&47:34:05.6&23.90&21.31   & 1.46  &0.25      &Sa      \nl
             &&  &           &c&09:55:19.07&47:34:07.7&12.20&20.90   & 1.29  &0.46      &SB6     \nl
             &&  &           &d&09:55:19.10&47:34:15.2&19.40&20.23   & 1.03  &1.75      &QSO, z=1.73\tablenotemark{j}  \nl
             &&  &           &e&09:55:19.26&47:33:47.5& 8.62&22.27   & 1.84  &1.00      &Sb      \nl
RXJ0955.5+4735&4.3&22& red
&E&09:55:30.20&47:35:39.0&$\ldots$&$>$23.01&$>$2.03&$\ldots$&?
\nl
             &&  &           &a&09:55:32.15&47:35:18.0&28.81&18.86   & 0.37  &$\ldots$  &G star? \nl
             &&  &           &b&09:55:31.94&47:35:13.3&31.15&20.47   & 1.01  &$\ldots$  &M star  \nl
RXJ0955.5+4753&7.7&18&FG can &a&09:55:33.49&47:53:17.9&15.90&21.48
& 1.66  &0.52      &Sb      \nl
             &&  &           &b&09:55:35.31&47:53:20.1& 6.40&15.23   &-0.83  &$\ldots$  &G star  \nl
RXJ0955.5+4756&3.0&21&AGN can&a&09:55:33.68&47:56:34.2&13.90&18.08
&-0.10  &$\ldots$  &F star  \nl
             &&  &           &b&09:55:34.58&47:56:35.5&15.30&19.74   & 0.56  &0.59      &SB6     \nl
             &&  &           &c&09:55:35.09&47:56:41.6&22.88&19.64   & 0.52  &0.17      &SB6     \nl
RXJ0955.6+4713&5.8&18&D      &a&09:55:41.50&47:13:01.1& 9.10&20.88
& 1.30  &1.32      &QSO     \nl
             &&  &           &b&09:55:42.74&47:13:16.3&16.00&20.58   & 1.18  &0.18      &SB6     \nl
RXJ0955.6+4721&2.3&23&D      &a&09:55:39.59&47:21:42.0&16.70&20.44
& 0.72  &0.00      &Sb      \nl
             &&  &         &b,R&09:55:40.24&47:21:54.7&15.70&19.65   & 0.41  &0.48      &QSO  \nl
             &&&&&&&&\multicolumn{4}{c}{CDS90-R324, FIRST J095540.2+472154} \nl
RXJ0955.6+4733&3.1&19&gal can&a&09:55:36.17&47:33:19.1&15.00&17.34
&-0.38  &0.16      &S0      \nl
             &&  &           &b&09:55:36.00&47:33:48.7&15.02&22.49   & 1.68  &0.57      &Sb,     \nl
             &&&&&&&&&&\multicolumn{2}{c}{ FIRST J095536.0+473349 }\nl
RXJ0955.6+4751&6.7&19&D      &a&09:55:37.67&47:51:28.0& 4.60&21.41
& 1.58  &0.89      &*SB6     \nl RXJ0955.7+4709&0.9&22&D
&a&09:55:46.40&47:09:32.8& 4.20&20.76   & 0.44  &1.02      &QSO
\nl
             &&  &           &b&09:55:44.94&47:09:46.1&17.50&22.04   & 0.96  &$\ldots$  &M star  \nl
             &&  &           &c&09:55:45.94&47:09:56.9&20.45&22.06   & 0.97  &0.30      &Sb      \nl
RXJ0955.7+4736&4.2&29&D      &a&09:55:48.58&47:36:59.0&14.70&20.66
& 1.07  &1.58      &QSO     \nl
             &&  &           &b&09:55:47.68&47:36:58.2&12.40&21.25   & 1.31  &0.32      &SB6     \nl
RXJ0955.8+4721&3.5&18& D     &V&09:55:50.32&47:21:11.6&
7.86&$>$23.01&$>$1.94&$\ldots$  &?        \nl
             &&  &           &a&09:55:51.64&47:20:48.0&24.53&20.94   & 1.11  &$\ldots$  &M star  \nl
             &&  &           &b&09:55:49.41&47:20:43.9&21.68&21.86   & 1.48  &0.47      &Sa?     \nl
RXJ0955.9+4726&2.3&24&D      &a&09:55:57.71&47:26:15.1&17.70&19.42
& 0.31  &0.26      &SB2     \nl
             &&  &           &b&09:55:56.65&47:26:22.9&16.70&19.96   & 0.53  &0.36      &Sa      \nl
             &&  &           &c&09:55:58.24&47:26:26.7& 4.30&21.70   & 1.23  &$\ldots$  &K star  \nl
             &&  &           &d&09:55:58.68&47:26:38.2& 8.70&21.05   & 0.97  &0.31      &SB6     \nl
RXJ0956.0+4708&5.9&22&D      &a&09:56:02.39&47:08:02.0&13.70&17.25
&-0.13  &$\ldots$  &G star  \nl
             &&  &           &b&09:56:03.67&47:08:11.3& 2.80&20.29   & 1.07  &1.82      &QSO     \nl
RXJ0956.0+4714&3.1&23&AGN can&a&09:56:01.28&47:14:20.0&10.50&18.12
&-0.07  &$\ldots$  &*K star \nl RXJ0956.0+4738&2.6&20&empty
&a&09:56:01.63&47:38:27.0& 7.90&20.94   & 0.98  &0.26      &SB1
\nl
             &&  &           &b&09:56:03.77&47:38:38.5&18.40&21.24   & 1.10  &0.69      &QSO     \nl
RXJ0956.1+4741&3.3&16&D      &a&09:56:07.41&47:41:23.4&14.30&20.31
& 0.83  &0.38      &Sb      \nl
             &&  &           &b&09:56:07.85&47:41:09.9&18.74&20.82   & 1.04  &0.21      &Sb      \nl
             &&  &           &c&09:56:09.00&47:41:09.0&17.12&20.53   & 0.92  &0.42      &SB6     \nl
             &&  &           &d&09:56:07.97&47:41:43.2&19.13&20.43   & 0.88  &0.32      &Sb      \nl
RXJ0956.2+4738&3.2&19&D      &a&09:56:15.77&47:38:44.7& 7.20&21.93
& 1.46  &0.49      &*Sb      \nl RXJ0956.3+4716&3.6&26&D
&a&09:56:16.16&47:16:49.1&20.00&19.98   & 0.73  &1.75      &*QSO,
z=1.69\tablenotemark{j}  \nl RXJ0956.4+4710&5.4&18&empty
&a&09:56:29.40&47:10:08.5& 8.20&21.45   & 1.50  &2.24      &QSO
\nl
             &&  &           &b&09:56:28.07&47:10:11.7& 7.80&20.88   & 1.27  &0.44      &SB2     \nl
RXJ0956.6+4717&11.1&14&M can &a&09:56:39.65&47:16:52.7&13.10&15.47
&-0.57  &$\ldots$  &*M star  \nl RXJ0956.6+4734&1.7&20&D
&a&09:56:37.90&47:34:50.3&12.30&20.83   & 0.75  &1.25      &*Sc
\nl RXJ0956.7+4729&9.5&33&sat
&a&09:56:45.69&47:28:55.1&30.00&$>$22.80&$>$0.88&0.09      &Sa?
\nl
             &&  &           &b&09:56:45.72&47:29:10.7& 8.60&19.90   & 1.12  &0.97      &SB4?    \nl
             &&  &           &c&09:56:48.01&47:29:41.5&30.00&19.81   & 1.09  &0.24      &Sc?     \nl
RXJ0956.8+4743&2.7&22&D      &a&09:56:52.31&47:43:35.0& 9.50&20.71
& 0.90  &0.49      &QSO     \nl
             &&  &           &b&09:56:50.09&47:43:50.7&20.40&21.02   & 1.02  &0.21      &Sa?     \nl
RXJ0956.9+4731&5.0&17&D      &a&09:56:55.63&47:31:08.2& 9.80&18.57
& 0.31  &0.34      &*SB6     \nl RXJ0956.9+4746&10.2 &17&D
&a&09:56:54.14&47:46:54.8&14.90&21.10   & 1.63  &0.48      &*SB2
\nl RXJ0956.9+4759&2.2&29&D
&a&09:56:55.42&47:58:43.1&28.50&18.14   &-0.21  &$\ldots$  &G star
\nl
             &&  &           &b&09:56:55.61&47:59:08.1& 4.90&18.76   & 0.03  &0.33      &Sa      \nl
             &&  &           &c&09:56:56.59&47:59:31.1&21.00&21.75   & 1.23  &$\ldots$  &M star  \nl
             &&  &           &d&09:56:58.09&47:59:30.8&28.80&20.52   & 0.74  &0.28      &Sb      \nl
RXJ0957.1+4727&2.3&16&D      &a&09:57:06.17&47:27:03.7&14.80&17.63
&-0.39  &$\ldots$  &M star  \nl
             &&  &           &b&09:57:07.91&47:27:09.7& 7.00&21.43   & 1.12  &0.77      &SB3     \nl
             &&  &           &c&09:57:06.95&47:27:21.5& 9.20&20.15   & 0.61  &0.07      &Sb      \nl
RXJ0957.2+4758&12.4&20&D     &a&09:57:13.00&47:58:38.0&19.20&19.61
& 1.12  &$\ldots$  &K star  \nl
             &&  &           &b&09:57:12.34&47:58:49.2& 8.60&20.11   & 1.32  &0.26      &SB4     \nl
             &&  &           &c&09:57:13.80&47:58:56.1&11.10&22.29   & 2.19  &$\ldots$  &M star  \nl
             &&  &           &d&09:57:14.48&47:58:42.2&23.20&20.31   & 1.41  &0.27      &SB2     \nl
             &&&&&&&&&&\multicolumn{2}{c}{b\&d grp?} \nl
RXJ0957.3+4729&2.6&21&D      &a&09:57:18.73&47:28:55.6& 6.50&21.20
& 1.08  &0.29      &SB6     \nl
             &&  &           &b&09:57:17.32&47:28:59.8&19.10&19.07   & 0.28  &0.16      &SB5, z=0.154\tablenotemark{j}  \nl
RXJ0957.3+4746&8.3&20&AGN can&a&09:57:22.56&47:46:04.3&17.78&22.63
& 2.16  &0.29      &SB5     \nl
             &&  &           &b&09:57:20.20&47:46:37.7&26.36&19.67   & 0.98  &2.25      &QSO     \nl
             &&  &           &V&09:57:21.37&47:46:04.4&19.94&$>$23.01&$>$2.31&$\ldots$  &?        \nl
RXJ0957.4+4746&8.2&20&AGN can&a&09:57:28.42&47:46:00.9&19.80&20.52
& 1.31  &$\ldots$  &M star  \nl
             &&  &           &b&09:57:30.07&47:46:09.3& 6.30&19.38   & 0.85  &1.76      &QSO     \nl
RXJ0957.5+4734&1.0&29&AGN can&a&09:57:35.52&47:34:49.5&23.30&21.89
& 0.94  &0.52      &Sa, \nl
             &&  &           &b&09:57:35.94&47:34:30.0&19.73&22.80   & 1.31  &0.40      &QSO     \nl
RXJ0957.5+4741&4.4&18&D      &a&09:57:32.65&47:41:52.1& 7.00&20.14
& 0.88  &0.81      &*QSO     \nl RXJ0957.6+4716&3.1&26&D
&a&09:57:36.57&47:16:25.2& 8.60&20.63   & 0.93  &0.52      &*SB6
\nl RXJ0957.6+4801&20.3&14&AGN   &a&09:57:38.51&48:01:21.7&
0.30&19.08   & 1.12  &1.11      &*QSO,    \nl
             &&&&&&&&&&\multicolumn{2}{c}{HS0954+4815, z=0.83\tablenotemark{j}} \nl
RXJ0957.7+4736&1.9&22&st can &a&09:57:43.72&47:36:09.8& 7.90&16.09
&-1.09  &$\ldots$  &F star  \nl
             &&  &           &b&09:57:44.78&47:36:12.8&14.60&20.04   & 0.48  &0.31      &SB6     \nl
             &&  &           &c&09:57:42.84&47:36:21.7& 7.40&16.74   &-0.83  &$\ldots$  &K star  \nl
RXJ0957.7+4745&30.8&12&AGN &a,R&09:57:46.59&47:45:49.0& 6.10&19.91
& 1.64  &0.74      &*QSO,     \nl
             &&&&&&&&\multicolumn{4}{c}{ CDS90-R346, FIRST J095746.6+474549,z=0.42\tablenotemark{j}}  \nl
RXJ0957.8+4713&3.1&27& D     &a&09:57:52.74&47:13:08.0&23.20&21.28
& 1.19  &$\ldots$  &M star  \nl
             &&  &           &b&09:57:50.41&47:13:12.5& 1.00&19.93   & 0.65  &0.58      &SB6     \nl
RXJ0957.8+4756&2.7&22& D     &a&09:57:53.27&47:56:08.2&12.80&19.11
& 0.26  &$\ldots$  &K star  \nl
             &&  &           &b&09:57:54.06&47:56:27.8&10.20&19.93   & 0.59  &1.78      &QSO     \nl
             &&  &           &c&09:57:53.64&47:56:16.8& 5.41&19.49   & 0.42  &$\ldots$  &M star  \nl
             &&  &           &d&09:57:51.13&47:56:23.7&21.97&21.06   & 1.04  &0.77      &Sa      \nl
RXJ0958.0+4728&2.8&22&empty  &a&09:58:03.57&47:28:44.6&17.20&20.87
& 0.98  &$\ldots$  &M star  \nl
             &&  &           &b&09:58:03.70&47:28:54.2&14.30&21.19   & 1.11  &0.33      &SB6     \nl
RXJ0958.0+4737&6.9&16&D      &a,R&09:58:00.63&47:37:34.8&
2.20&20.51   & 1.23  &0.50    &*SB5     \nl
             &&&&&&&&&&\multicolumn{2}{c}{   FIRST J095800.6+473734 } \nl
RXJ0958.0+4746&2.8&21&sat
&a&09:58:01.51&47:46:10.7&13.70&$<$11.80&$<$-2.64&$\ldots$ &*star
(saturated) \nl RXJ0958.2+4742&1.8&17&empty
&a&09:58:16.15&47:42:13.9& 8.20&21.63   & 1.09  &0.94      &*QSO
\nl RXJ0958.3+4725&48.8&12&AGN &a,R&09:58:19.70&47:25:07.6&
5.30&18.64   & 1.33  &1.82      &*QSO,    \nl
             &&&&&&&&\multicolumn{4}{c}{  CDS90-R351, FIRST J095819.7+472508, z=1.88\tablenotemark{j}}  \nl
RXJ0958.4+4739&2.2&16&unid   &a&09:58:27.72&47:39:52.6& 6.70&21.73
& 1.22  &0.41      &S0? \nl
             & & &           &b&09:58:28.29&47:39:59.5&15.20&19.53   & 0.60  &1.66      &QSO, z=1.67\tablenotemark{j}  \nl
RXJ0958.5+4738&32.7&13&AGN   &a&09:58:33.48&47:38:54.7& 4.80&18.44
& 1.08  &1.45      &*QSO, z=0.42\tablenotemark{j}  \nl
RXJ0958.5+4742&2.7&18&empty  &a&09:58:32.65&47:42:13.2&13.50&21.55
& 1.24  &0.72      &QSO     \nl
             &&  &           &b&09:58:32.27&47:42:32.9&17.30&21.18   & 1.09  &0.66      &SB1     \nl
RXJ0958.7+4800&2.2&21& D     &a&09:58:47.49&48:00:28.2&11.30&21.70
& 1.21  &0.69      &Sa      \nl
             &&  &           &b&09:58:48.99&48:00:09.6&21.67&21.15   & 0.99  &0.14      &Sb      \nl
RXJ0958.8+4715&2.0&24&FG can &a&09:58:54.21&47:15:06.8&21.10&19.93
& 0.46  &0.28      &SB6     \nl
             &&  &           &b&09:58:52.45&47:15:06.7& 5.90&13.80   &-1.98  &$\ldots$  &K star  \nl
RXJ0958.8+4719&2.8&17&red    &a&09:58:49.92&47:20:04.2& 7.90&17.49
&-0.36  &$\ldots$  &K star  \nl
             &&  &           &b&09:58:51.07&47:20:07.1&11.40&14.38   &-1.61  &$\ldots$  &K star  \nl
RXJ0958.8+4744&1.6&22&unid   &a&09:58:51.14&47:44:16.5&11.10&16.90
&-0.84  &0.01      &*HII galaxy, \nl
             &&&&&&&&&&\multicolumn{2}{c}{UGC 5354, z=0.0038\tablenotemark{j}}  \nl
RXJ0958.8+4801&1.5&24& D     &a&09:58:51.79&48:00:58.2&21.70&18.94
&-0.05  &0.10      &SB4     \nl
             &&  &           &b&09:58:52.89&48:01:01.1&10.30&21.59   & 1.00  &0.77      &Sa      \nl
             &&  &           &c&09:58:54.90&48:01:16.7&17.00&20.25   & 0.46  &1.95      &QSO     \nl
             &&  &           &d&09:58:52.90&48:01:21.1&20.70&20.69   & 0.64  &0.51      &SB6     \nl
             &&  &           &e&09:58:52.23&48:01:22.6&25.79&20.78   & 0.68  &0.26      &SB6     \nl
RXJ0958.9+4721&5.4&17& D     &a&09:58:53.98&47:20:59.3&15.10&21.20
& 1.40  &0.29      &Sb      \nl
             &&  &           &b&09:58:54.22&47:21:10.0& 4.90&19.88   & 0.87  &0.09      &SB1     \nl
RXJ0958.9+4745&29.3&20& star
&a&09:58:57.03&47:45:34.2&10.70&$<$9.90 &$<$-2.38&$\ldots$ &*F8
star\nl
             &&&&&&&&&&\multicolumn{2}{c}{BD+48-1823}
\tablenotetext{a}{Radial position uncertainty of X-ray detection,
in units of arcseconds.} \tablenotetext{b}{Preliminary
indentification based on HQS plates (Molthagen, Wendker \& Briel,
1997): D, visible on the HQS direct plate only; AGN=known AGN; AGN
can=AGN candidate (clssified as AGN or extremely blue); sat,
spectral type=star; galaxy.} \tablenotetext{c}{Object labels. i)
Sources with fluxes on more than a few passbands are labeled
a,b,c,$\ldots$ for association with each X-ray detection. ii) R =
known radio source, found by cross-comparing with the FIRST
catalog
(http://sundog.stsci.edu/first/catalogs/catalog\_01oct15.bin.gz),
NASA Extragalactic Database (NED) and SIMBAD; radio ID given in
last column according to FIRST and NED definitions. iii) V =
visible only in some direct images. iv) E = no source found in
x-ray error circle.} \tablenotetext{d}{Distance of the identified
optical source to the center of X-ray detection, in units of
arcseconds.} \tablenotetext{e}{V magnitude measure on our images.
If no detection, an upper limit is given; if saturated star, a
lower limit is given.} \tablenotetext{f}{The predicted redshift
for all non-stellar objects, as discussed in the text.}
\tablenotetext{g}{The ratio of X-ray to optical flux as defined in
the text.} \tablenotetext{h}{The new object classification based
on the SED information. ID's taken from NED and SIMBAD; see
Figure~2 and text for other definitions. An asterisk before this
classification indicates the most likely x-ray source.}
\tablenotetext{i}{Redshift as obtained from the following sources:
[RXJ0955.1+4729(e), RXJ0955.3+4733(d),
RXJ0956.3+4716(a),RXJ0957.3+4729(b), RXJ0957.6+4801(a),
RXJ0958.3+4725(a), RXJ0958.4+4739(b), RXJ0958.5+4738(a),
RXJ0958.8+4744(a); this paper, Fig.~6]; [RXJ0955.0+4720(a),
RXJ0957.7+4745(a); Mothagen, Wendker, \& Briel, 1997, A\&AS, 126,
509].}
\enddata
\end{deluxetable}